\documentclass[journal,comsoc]{IEEEtran}

\usepackage{graphicx}
\usepackage{epstopdf}
\usepackage{float}
\usepackage{algorithm,algorithmic}
\usepackage{array}
\usepackage{amsmath}
\usepackage{amssymb}
\usepackage{mdwmath}
\usepackage{mdwtab}
\usepackage{eqparbox}

\usepackage{fixltx2e}
\usepackage{cases}
\usepackage{bm}
\usepackage{multirow}
\usepackage{psfrag}
\usepackage[usenames]{color}
\usepackage{mathtools}

\usepackage{fixmath}
\usepackage{mathdots}
\usepackage{latexsym}
\usepackage{amsmath,amssymb}

\usepackage{bm}
\usepackage{footmisc}

\usepackage[T1]{fontenc}
\usepackage{url}
\usepackage{cite}
\newcommand{\subparagraph}{}
\usepackage{titlesec}

\newtheorem{theorem}{\bf Theorem}

\newcommand{\tabincell}[2]{\begin{tabular}{@{}#1@{}}#2\end{tabular}}

\begin{document}
\title{
	{Ordered Sequence Detection and Barrier Signal Design for Digital Pulse Interval Modulation in Optical Wireless Communications
}\\
}
\author{\IEEEauthorblockN{  Shuaishuai Guo,~\emph{Member, IEEE,}
Ki-Hong Park,~\emph{Member, IEEE},    
    and   Mohamed-Slim Alouini,~\emph{Fellow, IEEE}
                            }
\thanks{Part of the work has been presented in IEEE GLOBECOM OWC workshop 2018 \cite{Guo2018d}.}
\thanks{S. Guo, K.-H. Park and M. -S. Alouini are with King Abdullah University of Science and Technology (KAUST), Thuwal, Kingdom of Saudi Arabia, Saudi Arabia, 23955 (email: \{shuaishuai.guo; kihong.park; slim.alouini\}@kaust.edu.sa).}
\thanks{The work of Guo, K.-H. Park and M. -S. Alouini were supported by the funding from KAUST.
 }}
\maketitle
\begin{abstract}
This paper proposes an ordered sequence detection (OSD) for digital pulse interval modulation (DPIM) in optical wireless communications. Leveraging the sparsity of DPIM sequences, OSD shows comparable performance to the optimal maximum likelihood sequence detection (MLSD) with much lower complexity. Compared with the widely adopted sample-by-sample optimal threshold detection (OTD), it considerably improves the bit error rate (BER) performance by mitigating error propagation. Moreover, this paper proposes a barrier signal-aided digital pulse interval modulation (BDPIM), where the last of every $K$ symbols is allocated with more power as an inserted barrier signal.  BDPIM with OSD (BDPIM-OSD) can limit the error propagation between two adjacent barriers. To reduce the storing delay when using OSD to detect extremely large packets, we propose BDPIM with a combination of OTD and OSD (BDPIM-OTD-OSD), within which long sequences are cut into pieces and separately detected.
Approximate upper bounds of the average BER performance of DPIM-OTD, DPIM-OSD,  BDPIM-OSD and BDPIM-OTD-OSD are analyzed.
Simulations are conducted to corroborate our analysis.  Optimal parameter settings are also investigated in uncoded and coded systems by simulations. Simulation results show that the proposed
OSD and BDPIM bring significant improvement in  uncoded and coded systems over various channels. 
\end{abstract}
\begin{IEEEkeywords}
Digital pulse interval modulation, ordered sequence detection, signal design, error propagation, bit error rate
\end{IEEEkeywords}

\IEEEpeerreviewmaketitle

\section{Introduction}
\IEEEPARstart{D}{igital} pulse interval modulation (DPIM) \cite{Ghassemlooy2012} is an energy-efficient modulation technique for optical wireless communications based on intensity modulation and direct detection (IM/DD). Compared with conventional on-off keying (OOK), pulse amplitude modulation (PAM), pulse width modulation (PWM) and pulse position modulation (PPM), it utilizes pulse intervals to carry data and does not require accurate symbol-level synchronization.
Since proposed, it has been widely investigated in spectral evaluation \cite{Cariolaro2001,Ono2001} and error performance over various channels \cite{Zhang2000,Ghassemlooy2001,Ghassemlooy2003,Ghassemlooy2005,Popoola2006,Ghassemlooy2008,Ma2010,Jin2010,Jiang2014,Liu2014,Xu2015,Dabiri2016,Das2019}. Many enhanced modulation schemes have been proposed
\cite{Ghassemlooy2006,Aldibbiat2001,Ono2002,Ono2004,Sethakaset2005,Numata2011,He2013,Abdullah2014,Liao2014,Mi2016}. For instance, the authors of \cite{Ghassemlooy2006}  and \cite{Aldibbiat2001} proposed multilevel digital pulse interval modulation (MDPIM) and dual-header pulse interval modulation (DHPIM), which improved the information-carrying capability by additionally varying  pulse amplitudes and  pulse widths, respectively.

\begin{figure*}[t]
  \centering
    \includegraphics[width=0.9\textwidth]{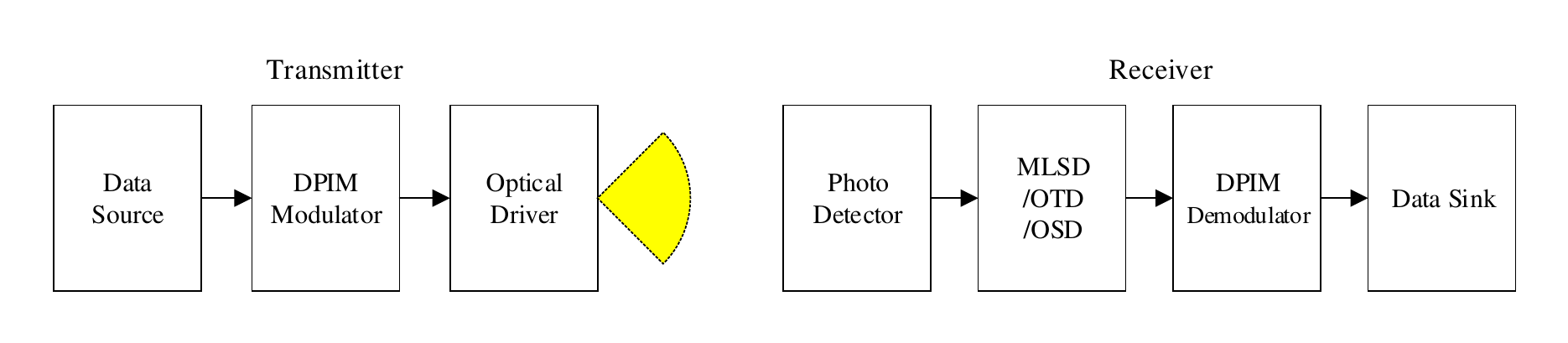}\\
  \caption{System model of DPIM}
  \label{System_Model}
\end{figure*}

DPIM symbols do not have equal duration. The fact complicates the analysis and application of DPIM.  Note that a single-chip error not only corrupts the bits directly associated with that chip but also shifts the bits that follow those bits \cite{Shiu1999}. Another distinct property resulting from the nonuniform symbol length is that the symbol boundaries are
not known before detection. Conventional soft decoding approaches for invariant symbol duration such as the Viterbi algorithm are not applicable \cite{Shiu1999}. The optimal soft decoding of
DPIM requires the use of high-complexity maximum likelihood sequence detector (MLSD). Therefore, most practical implementations of
DPIM would probably employ hard-decision decoding, i.e., sample-by-sample optimal threshold detection (OTD). Even though OTD has low complexity, it suffers from more severe bit error propagation, which leads to high bit error ratio (BER) in an erroneous packet. When using forward error correction (FEC) codes to correct the erroneous packet, lots of redundancy must be added in codewords because of the high BER thus resulting in a low rate. In other words, because of the high BER in an erroneous packet, DPIM can not benefit from the association with high-rate FEC codes for packet correction. To fix this problem, the error propagation of DPIM should be mitigated. 
Motivated by improving the system reliability, this paper proposes a low-complexity detection method that achieves the same performance as MLSD leveraging the sparsity of  DPIM sequences. Moreover, the paper proposes a barrier signal design for DPIM to limit the error propagation. It should be mentioned that
we adopt the average BER metric to measure the system reliability in this paper, which is defined by the packet error rate (PER) multiplying the expected BER in an erroneous packet.  Compared with the PER metric used in  \cite{Zhang2000,Ghassemlooy2001,Ghassemlooy2003,Ghassemlooy2005,Popoola2006,Ghassemlooy2008,Ma2010,Jin2010,Jiang2014,Liu2014,Xu2015,Dabiri2016,Das2019},
the average BER metric can measure both the PER and the error propagation level in an erroneous packet. The contributions of this paper are summarized as follows.
 \begin{itemize}
\item At the receiver side, the optimal MLSD of DPIM is formulated as a sparsity-constrained least square minimization problem. An ordered sequence detection (OSD) is proposed to solve the problem with low complexity.  The computational complexity and detection delay of OSD are analyzed. The relationship between OSD and the orthogonal matching pursuit (OMP) algorithm in \cite{Tropp2007} is revealed, which explains why OSD can achieve the same performance as MLSD.
\item At the transmitter side, a barrier signal-aided digital pulse interval modulation (BDPIM) is proposed to limit the error propagation and jointly utilized with OSD.  For large packet detection, we propose to use a combination of OTD and OSD to reduce the storage delay.  The parameter optimization of BDPIM is investigated by simulations.  

\item Approximate upper bounds of the average BER performance of DPIM with OTD (DPIM-OTD), DPIM with OSD (DPIM-OSD), BDPIM with OSD (BDPIM-OSD) and BDPIM with a combination of OTD and OSD (BDPIM-OTD-OSD) are derived.  Compared with \cite{Guo2018d}, the approximate BER upper bounds provided in this paper are applicable to all SNR regimes. Simulations are conducted to corroborate our theoretical analysis.

\item Comparisons are made among coded and uncoded  DPIM-OTD, DPIM-OSD, BDPIM-OSD,  BDPIM-OTD-OSD, etc. For extension, we investigate the performance evaluation over Gamma-Gamma turbulence channels. 
 \end{itemize}

 The remainder of the paper is organized as follows.  Section II describes the system model and the proposed designs. Section III presents the approximate upper bounds of the average BER performance of all described systems.
In Section IV, we show the superiority of the proposed schemes by simulations, investigate the impact of parameter settings on performance and discuss the extension to Gamma-Gamma  turbulence channels. Conclusions are drawn in the last section.

In this paper, the normal letter $x$ represents a scalar. The boldface letter $\textbf{x}$ stands for a vector. $||\textbf{x}||_0$ and $||\textbf{x}||_2$ denote the $l_0$ norm and $l_2$ norm of vector $\textbf{x}$, respectively. $\mathbb{R}$ represents the real-number field. $\mathbb{E}(\cdot)$ represents the expectation operation. $\lfloor x\rceil$ denotes the nearest integer to $x$. $\ln$ and $\log$ stand for the logarithm functions of base $e$ and base $2$, respectively. $\mathcal{Q}(\cdot)$ denotes the tail distribution function of the standard normal distribution. $\mathrm{erf}(\cdot)$ is the Gaussian error function. $n~\mathrm{mod}~K=0$ when $n$ is a multiple of $K$ and otherwise $n~\mathrm{mod}~K\neq0$.

\begin{figure}[t]
  \centering
 \includegraphics[width=0.42\textwidth]{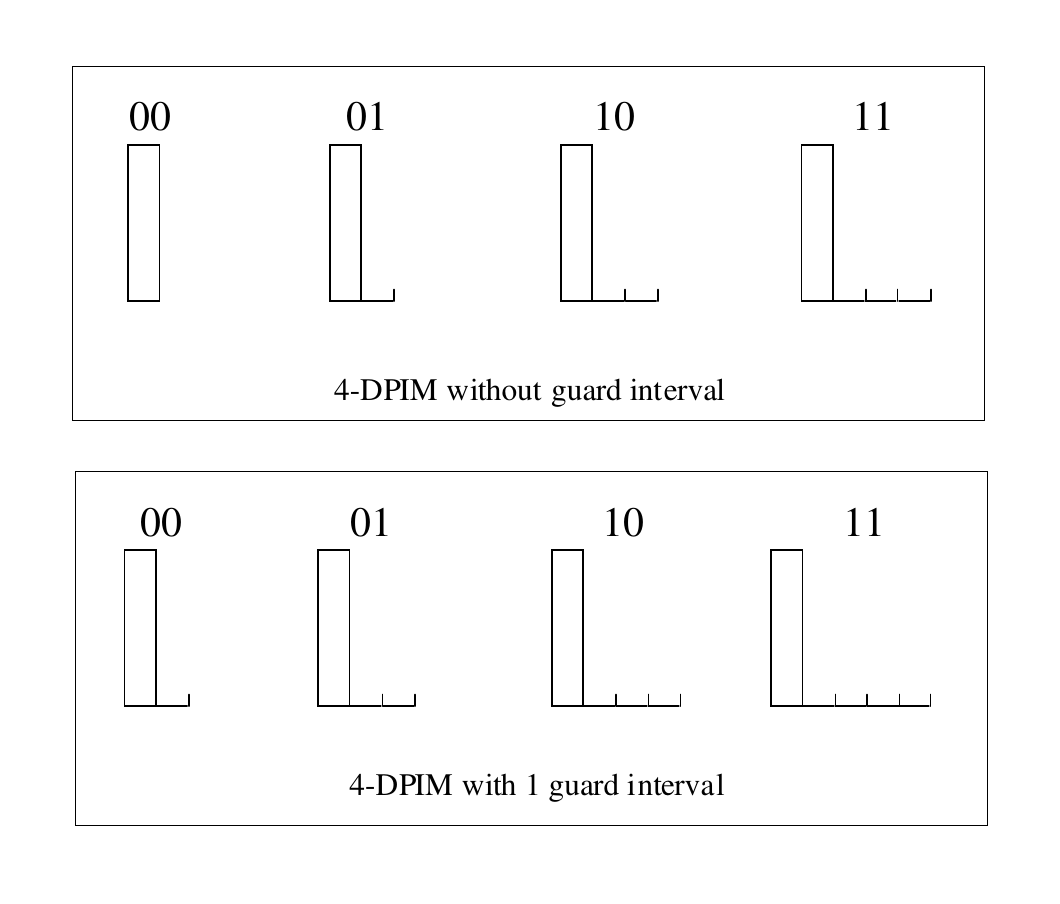}\\
  \caption{The signal forms of DPIM}
  \label{Signal_form}
  \end{figure}
  
\section{System Model and Proposed Designs}
In this section, the system model of conventional DPIM is first presented and then followed by an introduction to existing detection methods (i.e., MLSD, OTD) and the proposed OSD.  Also,   the new signal design, i.e., BDPIM is  described. 
\subsection{System Model of DPIM}
We consider a DPIM system as illustrated in Fig. \ref{System_Model}, where  $N_s$-symbol packets are transmitted by using $M$-ary DPIM.  The signal forms for DPIM has two types with and without guard intervals (GIs).
To demonstrate their differences,  we illustrate the signal forms of $4$-DPIM  in Fig. \ref{Signal_form}.  The data bits $\{00,~01,~10,~11\}$ are mapped to  $\{\mathrm{A},~ \mathrm{A}0,~\mathrm{A}00,~\mathrm{A}000\}$ with no GI, and to $\{\mathrm{A}0,~\mathrm{A}00,~\mathrm{A}000,~\mathrm{A}0000\}$ with $1$ GI, where $\mathrm{A}$ represents the peak amplitude and is proportional to the optical power. For the sake of simplicity, we assume that the proportional coefficient is $1$. In this paper, a general case with $g$ ($g\in\mathbb{N}$) GIs is considered
and the average symbol duration is given by
\begin{equation}
L_s=\frac{1}{M}\sum_{l=1}^{M}l+g=\frac{M+2g+1}{2}.
\end{equation}
Note that because the symbol duration of DPIM is variant, the total sequence length denoted by $L$ is also variant and the expected sequence length of $N_s$-symbol packets is $\mathbb{E}\{L\}=N_sL_s$.

 Let $h$ denote the channel coefficient.  The received signal vector $\textbf{y}\in\mathbb{R}^{L}$ can be written as
\begin{equation}\label{sig_model}
\textbf{y}=h\textbf{x}+\textbf{n},
\end{equation}
where $\textbf{x}\in\mathbb{R}^{L}$ represents the transmit signal vector, and $\textbf{n}\in\mathbb{R}^{L}$ is real-valued additive white Gaussian noise (AWGN) with zero mean and variance $\sigma_n^2\textbf{I}_L$. Moreover, the electrical SNR is denoted as $\gamma$ which is expressed as $\gamma=\mathrm{A}^2/\sigma_n^2$.
\subsection{Detection Methods}
\subsubsection{MLSD}
It is assumed that $h$ is known at the receiver. With full knowledge of $h$, $N_s$ and $L$, MLSD is given by
\begin{equation}
\begin{split}
(\textbf{P1}):\hat{\textbf{x}}&=\arg\max_{||\textbf{x}||_0=N_s}p_{\textbf{Y}}(\textbf{y}|\textbf{x},h)\\
&=\arg\min_{||\textbf{x}||_0=N_s}||\textbf{y}-h\textbf{x}||_2^2,
\end{split}
\end{equation}
since the probability density function (PDF) of \textbf{y} conditioned
on \textbf{x} and $h$ is
\begin{equation}
p_{\textbf{Y}}(\textbf{y}|\textbf{x},h) \propto \exp(-||\textbf{y}-h\textbf{x}||_2^2).
\end{equation}
There are a total of $\left(L\atop N_s\right)$ feasible solutions for  $\hat{\textbf{x}}$. For each feasible solution, $L$ real-number multiplications are needed to compute the $l_2$ norm. Therefore, MLSD based on exhaustive search requires $\left(L\atop N_s\right)L$ multiplications. Due to the high computational complexity,  it  is prohibitive especially for large packet detection.

\subsubsection{OTD} OTD detects the sequence using an optimal threshold $h\mathrm{A}_\mathrm{T}$. The key of OTD is finding $\mathrm{A}_\mathrm{T}$ based on the minimizing chip-error probability criterion. Specifically, the chip error probability can be expressed as
\begin{equation}
\begin{split}
P_c=&\mathrm{Pr}(0)\mathrm{Pr}(0\rightarrow \mathrm{A})+\mathrm{Pr}(\mathrm{A})\mathrm{Pr}(\mathrm{A} \rightarrow 0)\\
=&\mathrm{Pr}(0)\int_{\mathrm{A}_\mathrm{T}}^{\infty}\frac{h}{\sqrt{2\pi}\sigma_n}e^{-\frac{h^2x^2}{2\sigma_n^2}}dx
\\&~+\mathrm{Pr}(\mathrm{A})\int^{\mathrm{A}_\mathrm{T}}_{-\infty}\frac{h}{\sqrt{2\pi}\sigma_n}e^{-\frac{h^2(x-\mathrm{A})^2}{2\sigma_n^2}}dx.
\end{split}
\end{equation}
Taking the first derivative of $P_c$ with respect to $A_{T}$, we have
\begin{equation}
\frac{\partial P_c}{\partial\mathrm{A}_\mathrm{T}}=\mathrm{Pr}(\mathrm{A})\frac{h}{\sqrt{2\pi}\sigma_n}e^{-\frac{h^2(\mathrm{A}-\mathrm{A}_{\mathrm{T}})^2}{2\sigma_n^2}} -\mathrm{Pr}(0)\frac{h}{\sqrt{2\pi}\sigma_n}e^{-\frac{h^2\mathrm{A}_{\mathrm{T}}^2}{2\sigma_n^2}}.
\end{equation}
By solving the equation $\frac{\partial P_c}{\partial\mathrm{A}_\mathrm{T}}=0$, 
we obtain the optimal threshold $\mathrm{A}_{\mathrm{T}}$ as
\begin{equation}
\mathrm{A}_{\mathrm{T}}=\frac{\mathrm{A}}{2}-\frac{\mathrm{A}}{h^2\gamma}\ln\frac{\mathrm{Pr}(\mathrm{A})}{\mathrm{Pr}(0)}.
\end{equation}
From  (\ref{eqAt}), it is observed that $\mathrm{A}_{\mathrm{T}}$ is determined by $\mathrm{Pr}(\mathrm{A})$ and $\mathrm{Pr}(\mathrm{0})$, which are expressed as
\begin{equation}\label{PrA}
\mathrm{Pr}(\mathrm{A})=\frac{N_s}{L},
\end{equation}
and
\begin{equation}\label{Pr0}
\mathrm{Pr}(0)=\frac{L-N_s}{L},
\end{equation}
respectively. From (\ref{PrA}) and (\ref{Pr0}), it is observed that $N_s$ and $L$ should be be known for the determination of $\mathrm{A}_\mathrm{T}$. In practical implementation, the exact $\mathrm{Pr}(\mathrm{A})$, $\mathrm{Pr}(0)$ can be approximately given by their expectations which are $\mathbb{E}\{\mathrm{Pr}(\mathrm{A})\}=1/L_s$ and $\mathbb{E}\{\mathrm{Pr}(\mathrm{A})\}=1-1/L_s$. Based on these approximations, $\mathrm{A}_{\mathrm{T}}$ can be determined without the exact knowledge of $N_s$ and $L$ as
\begin{equation}\label{eqAt}
\mathrm{A}_{\mathrm{T}}\approx\frac{\mathrm{A}}{2}+\frac{ \mathrm{A}}{h^2\gamma}\ln{(L_s-1)}.
\end{equation}
\subsubsection{OSD} The optimal MLSD expressed as (\textbf{P1}) is indeed a sparsity-constrained least square minimization problem and the average sparsity level of DPIM sequences is $\mathbb{E}\{N_s/L\}=1/L_s$. According to \cite{Tropp2007}, the problem can be well solved by the OMP algorithm.
In use of the OMP algorithm,  the sequence can be iteratively detected as
\begin{equation}
\hat{t}_k=\arg \max_{t=1,2,\cdots,L} |\langle\textbf{r}_{k},\textbf{e}_t\rangle|
\end{equation}
with $\textbf{r}_1=\textbf{y}$ and  $\textbf{r}_{k}$ being updated by
\begin{equation}
\textbf{r}_{k+1}=\textbf{r}_{k}-\langle\textbf{r}_{k},\textbf{e}_{\hat{t}_k}\rangle \textbf{e}_{\hat{t}_k},
\end{equation}
where $\hat{t}_k$ is the position of $\mathrm{A}$ detected in the $k$th iteration, $\textbf{e}_t\in\mathbb{R}^L$ denotes the $t$th basis vector with $t$th entry being one and others being zeros.  It is observed from the procedure that OMP detects the $k$th largest receive signal as $\mathrm{A}$ in the $k$th iteration and repeats the detection until the number of iterations  equals to $N_s$.
 Instead of using the iterative procedure, we propose a simplified equivalent OSD in this paper. In OSD, a sorting algorithm is used to sort $\mathbf{y}=[y(1),y(2),\cdots,y(L)]$, the $N_s$ largest of which are detected as $\mathrm{A}$ and the others as $0$. Let $\{i(t)\}$ represent the order of $\{y(t)\}$, where $i(t)=1$ means that $y(t)$ is the largest. Mathematically,  the detected signal sequence can be written as
\begin{equation}
d(t)=\begin{cases}
\mathrm{A}, & 1\leq i(t) \leq N_s,\\
0, &N_s+1\leq i(t) \leq L.
\end{cases}
\end{equation}
Then by inputting $\mathbf{d}=[d(1),d(2),\cdots,d(L)]$ into conventional DPIM demodulator, we finally obtain the data bits. Moreover, it should be mentioned that $h$ is not used in the OSD procedure, which means that channel estimation is not required.
\begin{table*}[t]
\centering
\caption{Comparisons of Three Detection Methods}\label{tab1}
\footnotesize
    \begin{tabular}{ | c | c | c | c|c|c|}
    \hline
    Method & Detection manner & \tabincell{c}{Computation \\ complexity/delay} & Storage delay  &Performance & \tabincell{c}{Required information} \\ \hline
    MLSD & packet-by-packet &  $O\left(\left(L\atop N_s\right)L\right)$,  High & Low if $L$ is small &Optimal&$h$, $L$, $N_s$\\ \hline
   OTD & sample-by-sample &  $O\left(L\right)$, Low & Low&Suboptimal&$h$, $L_s$\\ \hline
   OSD & packet-by-packet &  $O\left(L\log L\right)$, Low  & Low if $L$ is small&Optimal&$L$, $N_s$\\ \hline
    \end{tabular}
\end{table*}
\begin{table*}[]
\caption{Data mapping of PPM, DPIM, MDPIM, DHPIM and BDPIM}
\center
\begin{tabular}{lllllll}
Data& PPM & DPIM & MDPIM\cite{Ghassemlooy2006} & $\textrm{DHPIM}$\cite{Aldibbiat2001}&  $\textrm{BDPIM}_{n~\mathrm{mod}~K\neq0}$&$\textrm{BDPIM}_{n~\mathrm{mod}~K=0}$\\
 000&$\mathrm{A}$0000000&$\mathrm{A}$0&$\mathrm{A}_{\mathrm{L}}0$ &$\mathrm{A}$0& $\mathrm{A}_{\mathrm{L}}$0&$\mathrm{A}_{\mathrm{H}}$0\\
 001&0$\mathrm{A}$000000&$\mathrm{A}$00 &$\mathrm{A}_{\mathrm{L}}$00  &$\mathrm{A}$00 &$\mathrm{A}_{\mathrm{L}}$00&$\mathrm{A}_{\mathrm{H}}$00\\
 010&00$\mathrm{A}$00000&$\mathrm{A}$000&$\mathrm{A}_{\mathrm{L}}$000& $\mathrm{A}$000&$\mathrm{A}_{\mathrm{L}}$000&$\mathrm{A}_{\mathrm{H}}$000\\
 011&000$\mathrm{A}$0000&$\mathrm{A}$0000&$\mathrm{A}_{\mathrm{L}}$0000&$\mathrm{A}$0000&$\mathrm{A}_{\mathrm{L}}$0000&$\mathrm{A}_{\mathrm{H}}$0000\\
100&0000$\mathrm{A}$000&$\mathrm{A}$00000&$\mathrm{A}_{\mathrm{H}}$0&$\mathrm{A}$$\mathrm{A}$0& $\mathrm{A}_{\mathrm{L}}$00000&$\mathrm{A}_{\mathrm{H}}$00000\\
 101&00000$\mathrm{A}$00&$\mathrm{A}$000000 &$\mathrm{A}_{\mathrm{H}}$00  &$\mathrm{A}$$\mathrm{A}$00 &$\mathrm{A}_{\mathrm{L}}$000000&$\mathrm{A}_{\mathrm{H}}$000000\\
 110&000000$\mathrm{A}$0&$\mathrm{A}$0000000& $\mathrm{A}_{\mathrm{H}}$000 &$\mathrm{A}$$\mathrm{A}$000 &$\mathrm{A}_{\mathrm{L}}$0000000&$\mathrm{A}_{\mathrm{H}}$0000000\\
 111&0000000$\mathrm{A}$&$\mathrm{A}$00000000&$\mathrm{A}_{\mathrm{H}}$0000&$\mathrm{A}$$\mathrm{A}$0000  &$\mathrm{A}_{\mathrm{L}}$00000000&$\mathrm{A}_{\mathrm{H}}$00000000\\
\end{tabular}
\end{table*}

\emph{Computational Complexity and Delay Analysis of Three Detection Methods:} As analyzed in Section II-B-1,  MLSD based on exhaustive search requires $\left(L\atop N_s\right)L$ multiplications, which can be expressed as $\mathcal{O}\left(\left(L\atop N_s\right)L\right)$.
OTD requires $L$ comparisons, the complexity of which can be given by $\mathcal{O}\left(L\right)$. OSD can sort $L$ received signals by using the QuickSort algorithm, whose complexity is $\mathcal{O}(L\log L)$.  
The detection delay includes the storage latency and the computation delay. Since MLSD and OSD are operated packet-by-packet, the storage delay of an $L$-chip packet is
\begin{equation}\label{tau1}
\tau_{1}=\frac{L}{R_{c}},
\end{equation}
where $R_{c}$ is the transmission rate in chips per second. Differently, OTD is performed sample-by-sample, and thus the storage delay of OTD is
\begin{equation}
\tau_{2}=\frac{1}{R_{c}}.
\end{equation}
For high-rate optical wireless communications, the storage latency of OTD only depends on the transmission rate and thus it is low regardless of the sequence length. The latency of OSD and that of MLSD are also dependent on $L$, and therefore they are low only if $L$ is small.  The computational delay is proportional to the computational complexity. Since the complexity of OTD and OSD is quite low, the computation delay can be omitted. On the contrary, the complexity of MLSD is high thus resulting in a computation delay of $O\left(\left(L\atop N_s\right)L\right)$ unit times (UTs), where a UT is the amount of time it takes for a multiplication. For clearly viewing the differences of these detection methods, we list the comparison results in terms of complexity, storage delay, performance as well as the required information in Table I.

\subsection{Barrier Signal Design}
\begin{figure}[t]
  \centering
    \includegraphics[width=0.4\textwidth]{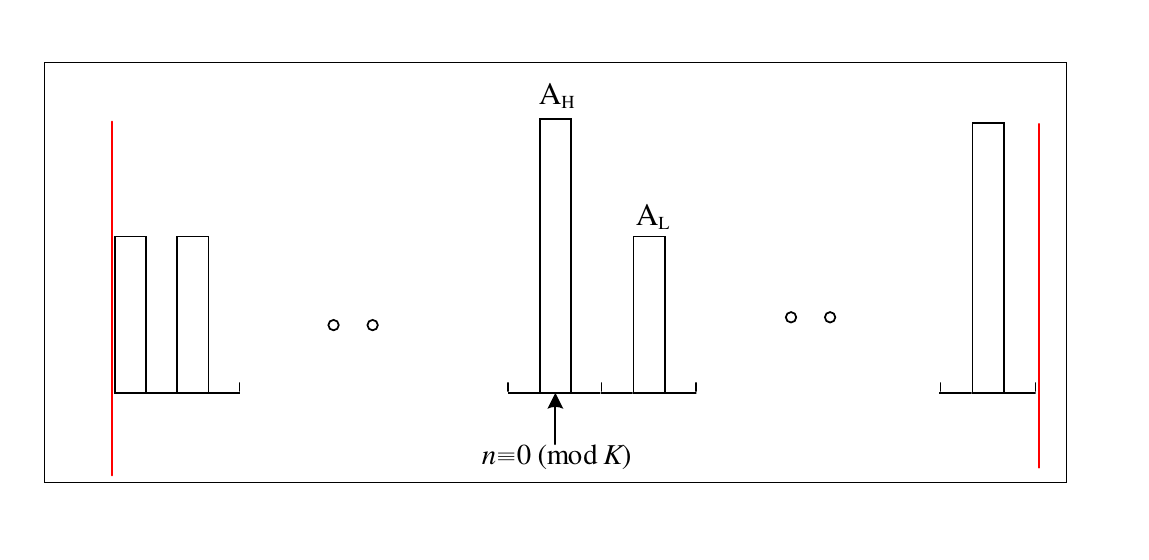}\\
  \caption{The signal forms of BDPIM}
  \label{Packet}
\end{figure}

The severe error propagation of DPIM mainly results from unknown symbol boundaries. To overcome it, we propose a barrier signal design, i.e., BDPIM.  As shown in  Fig. \ref{Packet}, BDPIM signals have two amplitudes. The amplitude of the $n$th symbol is $\mathrm{A}_{\mathrm{H}}$, when $n~\mathrm{mod}~K=0$  and otherwise $\mathrm{A}_{\mathrm{L}}$ in BDPIM, where $\mathrm{A}_{\mathrm{H}}>\mathrm{A}_{\mathrm{L}}$ and $K$ is a positive integer.  For fair comparison with DPIM under the same optical power constraint, the power allocation in every $K$ symbols of BDPIM  satisfies 
\begin{equation}\label{Con1}
(K-1)\mathrm{A}_{\mathrm{L}}+\mathrm{A}_{\mathrm{H}}=K\mathrm{A}.
\end{equation}
 To show the difference of the proposed design from existing signal designs, we list the data mapping of PPM, DPIM with 1 GI, MDPIM \cite{Ghassemlooy2006}, DHPIM\cite{Aldibbiat2001}, the proposed BDPIM in Table II. It is shown that BDPIM  has the same average symbol duration as DPIM, which is larger than that of MDPIM \cite{Ghassemlooy2006} and that of DHPIM \cite{Aldibbiat2001}.
For simplicity, we assume that the number of symbols $N_s$ in a packet can be divided by $K$, which means $N_s$ can be expressed as $N_s=KQ$ and $Q$ is a positive integer.

At the receiver, OSD for BDPIM has two phases. In the first OSD phase, $\mathbf{y}$ is sorted and the $Q$ largest are detected as $\mathrm{A}_{\mathrm{H}}$, the positions of which is recorded as $t_1, t_2,\cdots, t_Q$.  Let $t_1'< t_2' <\cdots< t_Q'$ denote the sorted positions in an ascent order and $t'_0=0$. In the second OSD phase,  $\mathbf{y}(t_i'+1:t_{i+1}'-1)$ is sorted, the $(K-1)$ largest values of which are detected as $\mathrm{A}_{\mathrm{L}}$ and others as $0$ for any $i=0,1,\cdots,Q-1$.  At last, by inputting the detected signal vector into conventional DPIM demodulator, we finally obtain the data bits.  The detailed OSD algorithm for BDPIM is listed in Algorithm 1.
\begin{algorithm}[t] 
\caption{OSD Algorithm for BDPIM}
\label{alg:osd}
\begin{algorithmic} 
\STATE  \textbf{Input:} $\mathbf{y}$, $L$, $N_s$ and $Q$
\STATE Sort $\mathbf{y}$ and output the order $\{i(t)\}$.
\STATE Detect $1\leq i(t)\leq Q$ as $\mathrm{A}_{\mathrm{H}}$ and output the indices $t_1, t_2,\cdots, t_Q$.
\STATE Sort $t_1, t_2,\cdots, t_Q$ in an ascent order $t_1'< t_2' <\cdots< t_Q'$.
\STATE $t_0=0$.
\FOR{$i=0:Q-1$}
\STATE Sort $\mathbf{y}(t_i'+1:t_{i+1}'-1)$ and detect the $(K-1)$ largest as $\mathrm{A}_{\mathrm{L}}$ and others as $0$.
\ENDFOR
\end{algorithmic}
 \end{algorithm}

As described, OSD for BDPIM includes two phases. The first involves a sorting algorithm operating on $L$ received signals, whose complexity is $\mathcal{O}(L\log L)$. The second involves sorting algorithms operating on $\mathbf{y}(t_i'+1:t_{i+1}'-1)$ for all $i=0,1,\cdots,Q-1$. The average length of 
 $\mathbf{y}(t_i'+1:t_{i+1}'-1)$ is around $L/Q$ and the sorting operations are conducted for $Q$ times. Therefore,  the complexity of the second detection phase is around $\mathcal{O}\left(L\log(L/Q)\right)$. In summary, the total complexity of OSD for BDPIM is also around $\mathcal{O}(L\log L)$.  
The storage delay is in the same order of OSD for DPIM and dependent on the packet size. For large packet detection, the delay may cause unbearable latency. To address this issue, we propose to use a combination of OTD and OSD. OTD is first used to detect $\mathrm{A}_{\mathrm{H}}$ according to an optimal threshold $h\mathrm{A}_{\mathrm{T}}'$. If a sample is not detected as $\mathrm{A}_{\mathrm{H}}$, we store it in a buffer. If it is, we activate OSD to detect the signal sequence in the buffer. In this way, long sequences are split into small pieces. Specifically, the sequence of a packet is divided into the pieces of average length $L/Q$.
Therefore, both the storage delay and  processing time can be reduced  by a factor of $Q$. 
\section{Main Results on BER Performance}
This section presents the main results on the approximate upper bounds of the average BER performance of all schemes, including DPIM-OTD, DPIM-OSD, BDPIM-OSD and BDPIM-OTD-OSD, where the average BER performance is defined by
\begin{equation}
P_e=\overline{\mathsf{BER}}P_p,
\end{equation}
where $\overline{\mathsf{BER}}$ denotes the expected BER in an erroneous packet reflecting the error propagation level and $P_p$ represents the PER. 
\subsection{Approximate BER Upper Bounds of DPIM-OTD}
 The PER of OTD can be written as
\begin{equation}\label{Pp1}
P_{p_1}=1-(1-P_c)^L,
\end{equation}
based on which and the approximate expression of $\mathrm{A}_{\mathrm{T}}$ in (\ref{eqAt}) we have the following theorem:
\begin{theorem} 
The average BER of DPIM-OTD
is approximately upper bounded by
\begin{equation}\label{Pe1}
\begin{split}
 P_{e_1}^\mathcal{U}
\approx\frac{2-2(1-P_c)^L-LP_c(1-P_c)^{L-1}}{4},
\end{split}
\end{equation}
where the chip error probability is approximated as
\begin{equation}\label{Pc}
\begin{split}
P_c&\approx\frac{L_s-1}{L_s} \mathcal{Q}\left(\frac{h\sqrt{\gamma}}{2}+\frac{1}{h\sqrt{\gamma}}\ln(L_s-1)\right)
\\&~~~~~+\frac{1}{L_s} \mathcal{Q}\left(\frac{h\sqrt{\gamma}}{2}-\frac{1}{h\sqrt{\gamma}}\ln(L_s-1)\right).
\end{split}
\end{equation}
\end{theorem}
\begin{IEEEproof}
See Appendix A.
\end{IEEEproof}
\emph{Remark and Observation:} From (\ref{Pe1}) and (\ref{Pc}), it is observed that $P_{e_1}^\mathcal{U}$ is a function of $\gamma$, $L$, $h$ and $L_s$.
Based on the  following inequalities of derivatives
\begin{equation}
\frac{\partial P_{e_1}^\mathcal{U}}{\partial P_c}>0,
\end{equation}
\begin{equation}
\frac{\partial P_c}{\partial\gamma}<0,
\end{equation}
and the chain rule, we have 
\begin{equation}
\frac{\partial P_{e_1}^\mathcal{U}}{\partial \gamma}=\frac{\partial P_{e_1}^\mathcal{U}}{\partial P_c}\cdot\frac{\partial P_c}{\partial\gamma}<0.
\end{equation}
This indicates that $P_{e_1}^\mathcal{U}$ decreases as $\gamma$ increases. 
Taking the first derivative of $P_{e_1}^\mathcal{U}$ with respect to $L$, we obtain
\begin{equation}
\begin{split}
\frac{\partial P_e^\mathcal{U}}{\partial L}\geq0,
\end{split}
\end{equation}
which shows that $P_{e_1}^\mathcal{U}$ increases as $L$ increases.

\begin{figure}[t]
  \centering
  \includegraphics[width=0.45\textwidth]{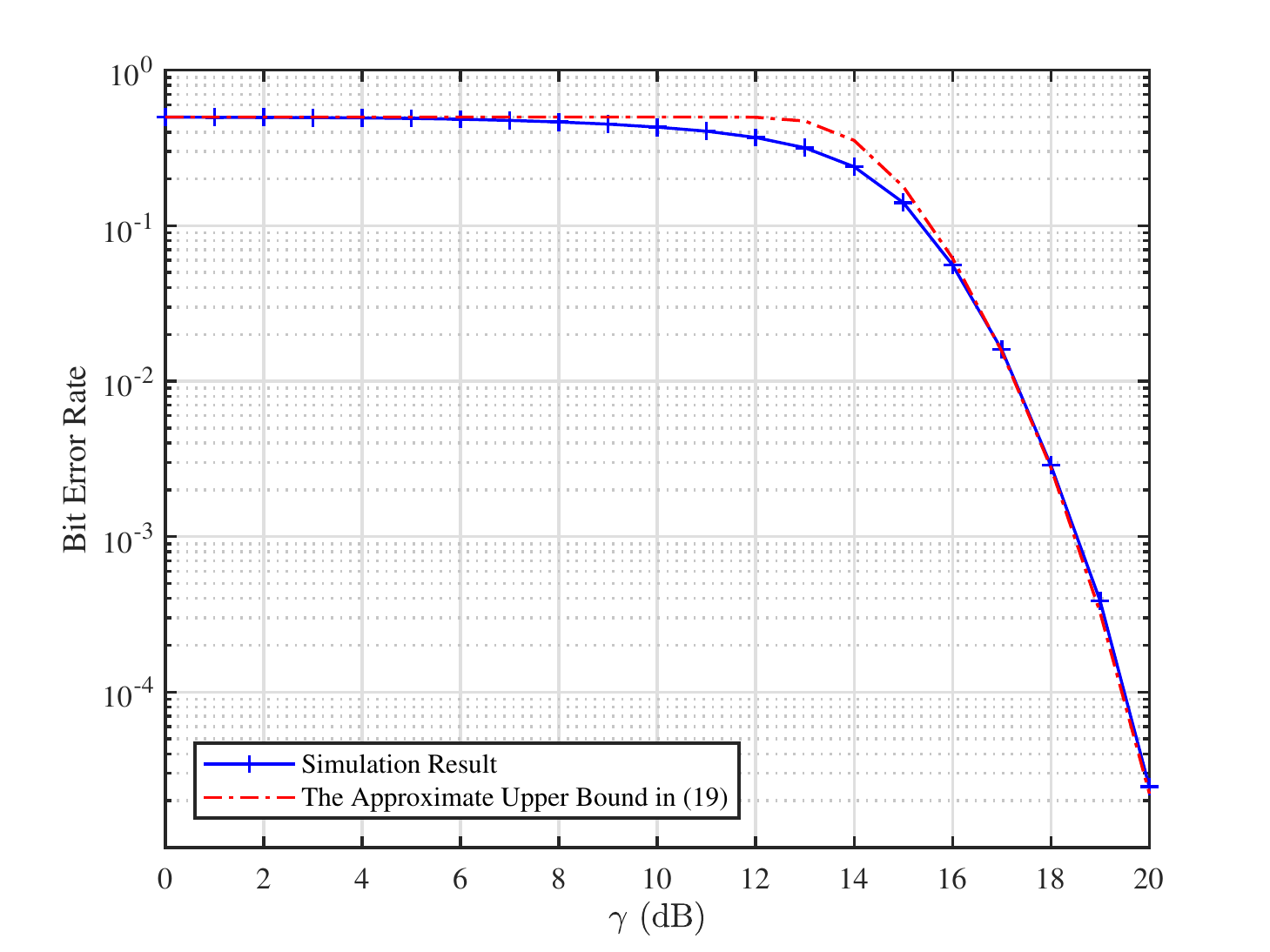}\\
 \caption{Approximate BER upper bound of DPIM-OTD where $h=1$, $N_s=100$ and $4$-DPIM with $1$-GI are employed.}
  \label{2a1}
\end{figure}

We demonstrate the simulation result of the average BER performance and the derived approximate upper bound of DPIM-OTD in Fig. \ref{2a1}. In the simulations, $h=1$, $N_s=100$ and $4$-DPIM with $1$ GI are used. 
The results are averaged over $10^5$ Monte Carlo simulations. Results show that the derived approximate upper bound matches the simulation result well especially in the low and high SNR regime. 

\subsection{Approximate BER Upper Bounds of DPIM-OSD}
In use of OSD, chip errors occur in pairs in an erroneous packet. In detail, as long as a \emph{false alarm error} occurs,  an \emph{erasure error} occurs, because the number of $\mathrm{A}$s by OSD is strictly equal to $N_s$. Based on the fact, the PER of DPIM-OSD is indeed the probability of there being at least one pair of chip errors and can be written as
\begin{equation}\label{eqPp2}
P_{p_2}=\mathrm{Pr}\{U_{1:L-N_s}>V_{N_s:N_s}\},
\end{equation}
where $U_{1:L-N_s}$ denotes the $1$st largest of $L-N_s$ received signals when $0$ is transmitted and $V_{N_s:N_s}$ denotes the $N_s$th largest (i.e., the smallest) of $N_s$ received signals when $\mathrm{A}$ is transmitted.  
For any random variables $U$ and $V$, the probability of $U>V$ can be calculated by
\begin{equation}
\begin{split}
\mathrm{Pr}\{U>V\}
&=\int_{-\infty}^{+\infty}[1-F_U(v)]f_V(v) dv,\\
\end{split}
\end{equation}
where $F_U(u)$ is the cumulative distribution function (CDF) of variable $U$ and  $f_V(v)$ is the PDF of variable $V$.
Given $U_{k_1:n_1}$ as the $k_1$th largest of $n_1$ real Gaussian variables with mean $\mu_1$ and variance $\sigma_n^2/h^2$,  $V_{k_2:n_2}$ as the $k_2$th largest of $n_2$ real Gaussian variables with mean $\mu_2$ and variance $\sigma_n^2/h^2$, we define an $\mathcal{OR}$ function to represent $\mathrm{Pr}\{U_{k_1:n_1}>V_{k_2:n_2}\}$ as
\setlength{\arraycolsep}{0.1pt}
\begin{equation}\label{ORF}
\begin{split}
&\mathcal{OR}{\scriptsize\left|\begin{array}{ccc}
\mu_1, & k_1, & n_1 \\
\mu_2,&k_2, & n_2 \\
\end{array}\right.}\triangleq\mathrm{Pr}\{U_{k_1:n_1}>V_{k_2:n_2}\}\\
& =\int_{-\infty}^{+\infty}[1-F_{U_{k_1:n_1}}(v)]f_{V_{k_2:n_2}}(v) dv.\\
\end{split}
\end{equation}
According to \cite{Yang2011},  the CDF $F_{U_{k_1:n_1}}(v)$ in (\ref{ORF}) is given by
\begin{equation}
\begin{split}
F_{U_{k_1:n_1}}(v)=\sum_{k=n_1+1-k_1}^{n_1}\left(n_1\atop k\right)F_U(v)^k(1-F_U(v))^{n_1-k},
\end{split}
\end{equation}
where
\begin{equation}
F_U(v)=\frac{1}{2}\left[1+\mathrm{erf}\left(\frac{hv-h\mu_1}{\sqrt{2}\sigma_n}\right)\right].
\end{equation}
The PDF $f_{V_{k_2:n_2}}(v)$ is written as 
\begin{equation}
\begin{split}
&f_{V_{k_2:n_2}}(v)=\\
&\frac{n_2!}{(n_2-k_2)!(k_2-1)!}F_V(v)^{n_2-k_2}[1-F_V(v)]^{k_2-1}f_{V}(v),
\end{split}
\end{equation}
where
\begin{equation}
f_V(v)=\frac{h}{\sqrt{2\pi}\sigma_n}e^{-\frac{h^2(v-\mu_2)^2}{2\sigma_n^2}},
\end{equation}
and 
\begin{equation}
F_V(v)=\frac{1}{2}\left[1+\mathrm{erf}\left(\frac{hv-h\mu_2}{\sqrt{2}\sigma_n}\right)\right].
\end{equation}

 Based on the definition of the $\mathcal{OR}$ function,
the PER of DPIM-OSD can be given by
\begin{equation}\label{Pp2}
P_{p_2}=\mathcal{OR}{\scriptsize\left|\begin{array}{ccc}
0, & 1, & L-N_s \\
\mathrm{A},&N_s, & N_s \\
\end{array}\right.},
\end{equation}
according to which, we derive the following theorem:
\begin{theorem}
The average BER of DPIM-OSD is approximately upper bounded by
\begin{equation}\label{Pe2a}
\begin{split}
 P_{e_2}^\mathcal{U}\approx\frac{1}{6}\left(\mathcal{OR}{\scriptsize\left|\begin{array}{ccc}
0,  & 1, & L-N_s \\
\mathrm{A},&N_s, & N_s \\
\end{array}\right.}+2\mathcal{OR}{\scriptsize\left|\begin{array}{cccc}
0, & 2, & L-N_s \\
\mathrm{A},&N_s-1, & N_s \\
\end{array}\right.}\right),
\end{split}
\end{equation}
and the approximation holds when $L$ is large.
\end{theorem}
\begin{IEEEproof}
See Appendix B.
\end{IEEEproof}

\emph{Remark and Observation:} The approximate upper bound ${P_{e_2}^\mathcal{U}}$ in (\ref{Pe2a}) is a function of  $h$, $\mathrm{A}$, $\sigma_n$, $L$ and $N_s$. It has more theoretical significance than practical significance since the defined $\mathcal{OR}$ function involves complicated integrals thus offering no clear insights into the system performance. To simplify the analytical result, we  propose an approximation of the $\mathcal{OR}$ function as
\begin{equation}
\begin{split}
\mathcal{OR}{\scriptsize\left|\begin{array}{ccc}
\mu_1,& k_1, & n_1 \\
\mu_2,&k_2, & n_2 \\
\end{array}\right.}&\approx\mathrm{Pr}\{U_{k_1:n_1}>v_{k_2:n_2}\}\\&=1-F_{U_{k_1:n_1}}(v_{k_2:n_2}),
\end{split}
\end{equation}
where $v_{k_2:n_2}=\alpha\mathbb{E}(V_{k_2:n_2})$ and $\alpha$ is a constant. We use $\alpha=0.82$ in this paper and the accuracy of the approximation  is investigated by simulations.  Based on the approximation, we have 
\begin{equation}
\begin{split}
\mathcal{OR}{\scriptsize\left|\begin{array}{ccc}
0,  & 1, & L-N_s \\
\mathrm{A},&N_s, & N_s \\
\end{array}\right.}&\approx 1-F_U(v_1)^{L-N_s},
\end{split}
\end{equation}
and
\begin{equation}
\begin{split}
&\mathcal{OR}{\scriptsize\left|\begin{array}{ccc}
0,  & 2, & L-N_s \\
\mathrm{A}, &N_s-1, & N_s \\
\end{array}\right.}
\approx 1-F_{U_{2:L-N_s}}(v_2)
\approx 1-F_{U_{2:L-N_s}}(v_1)
\\&=1-F_U(v_1)^{L-N_s}-
(L-N_s)F_U(v_1)^{L-N_s-1}[1-F_U(v_1)],
\end{split}
\end{equation}
where $v_{1}=\alpha\mathbb{E}(V_{N_s:N_s})$ and $v_2=\alpha\mathbb{E}(V_{N_s-1:N_s})$. Based on these approximations and Theorem 2, we deduce a tractable approximate BER upper bound as
\begin{equation}\label{Pe2b}
\begin{split}
 \overline{P_{e_2}^\mathcal{U}}\approx&\frac{1}{2}\left[1-F_U(v_1)^{L-N_s}\right]\\&-\frac{1}{3}(L-N_s)F_U(v_1)^{L-N_s-1}[1-F_U(v_1)].
\end{split}
\end{equation}
 The new approximate upper bound $\overline{P_{e_2}^\mathcal{U}}$ is a function of 
$h$, $\mathrm{A}$, $v_1$, $\sigma_n$, $L$ and $N_s$. It is noted that $\mathbb{E}(V_{N_s:N_s})$ in the expression of $v_1$ is the expectation of the least order statistics of $N_s$ Gaussian samples with mean $\mathrm{A}$ and covariance $\sigma_n^2/h^2$, which can be approximately computed by \cite{Chen1999}
\begin{equation}
\mathbb{E}(V_{N_s:N_s})=\mathrm{A}-\frac{\sigma_n\phi^{-1}\left(0.5264^{1/N_s}\right)}{h},
\end{equation}
where $\phi^{-1}(\cdot)$ is the inverse of the Gaussian CDF \cite{Chen1999}. $\overline{P_{e_2}^\mathcal{U}}$ is much simpler than (\ref{Pe2a}) and can be regarded as a closed-form expression because it only involves the $\mathrm{erf}$ function, which can be further approximated by \cite{Chiani2003}
\begin{equation}
\mathrm{erf}(v)\approx1-\frac{1}{6}e^{-v^2}-\frac{1}{2}e^{-\frac{4}{3}v^2}.
\end{equation}

\begin{figure}[t]
  \centering
  \includegraphics[width=0.45\textwidth]{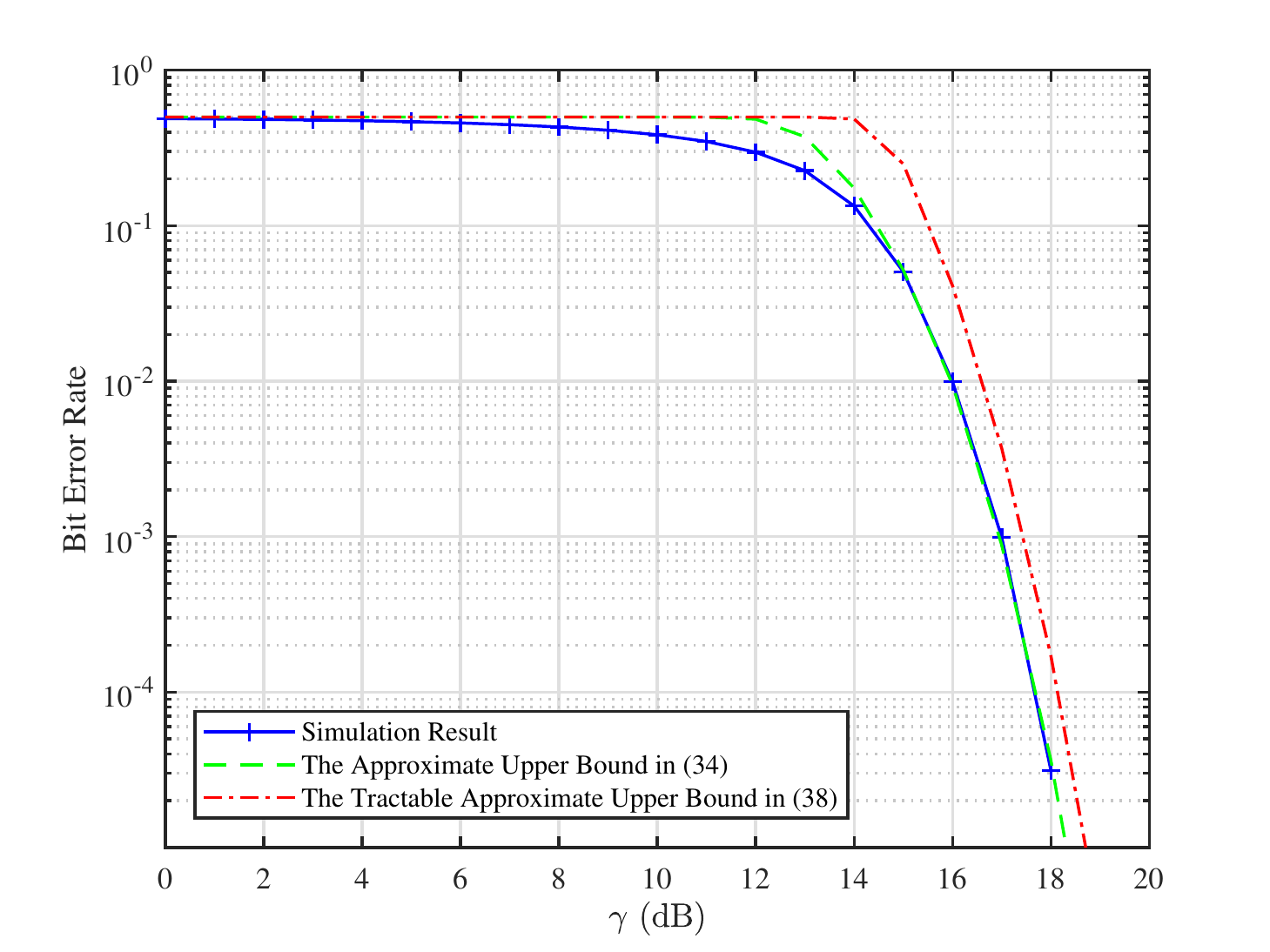}\\
 \caption{Approximate BER upper bounds of DPIM-OSD where $h=1$, $N_s=100$ and $4$-DPIM with $1$-GI are employed.}
  \label{2a2}
\end{figure}

Under the same simulation setups as that in Section III-A, we simulate the BER performance and compute the derived bounds numerically as illustrated in Fig. \ref{2a2}.  Results demonstrate that the approximate upper bound in (\ref{Pe2a}) matches the BER performance well, but is too complicated to analyze.  The tractable approximate upper bound in (\ref{Pe2b}) has a gap to the simulation result, but it enjoys low complexity.

\begin{figure*}
\setcounter{equation}{42}
\begin{equation}\label{Pe3a}
\begin{split}
 P_{e_3}^\mathcal{U}
\approx&\frac{1}{6}\left(\mathcal{OR}{\scriptsize\left|\begin{array}{cccc}
\mathrm{A}_{\mathrm{L}}, &  1, & N_s-Q \\
\mathrm{A}_{\mathrm{H}},& Q, & Q \\
\end{array}\right.}+2\mathcal{OR}{\scriptsize\left|\begin{array}{cccc}
\mathrm{A}_{\mathrm{L}}, &  2, & N_s-Q \\
\mathrm{A}_{\mathrm{H}},& Q-1, & Q \\
\end{array}\right.}\right)\left(1-\mathcal{OR}{\scriptsize\left|\begin{array}{ccc}
0, & 1, & \lfloor KL_s\rceil-K\\
\mathrm{A}_{\mathrm{L}},&K-1, &K-1 \\
\end{array}\right.}\right)\\
&+\frac{1}{6}\left(\mathcal{OR}{\scriptsize\left|\begin{array}{ccc}
0, & 1, & \lfloor KL_s\rceil-K\\
\mathrm{A}_{\mathrm{L}},&K-1, &K-1 \\
\end{array}\right.}+2\mathcal{OR}{\scriptsize\left|\begin{array}{ccc}
0, & 2, & \lfloor KL_s\rceil-K\\
\mathrm{A}_{\mathrm{L}},&K-2, &K-1 \\
\end{array}\right.}\right)\left(1-\mathcal{OR}{\scriptsize\left|\begin{array}{cccc}
\mathrm{A}_{\mathrm{L}}, &  1, & N_s-Q \\
\mathrm{A}_{\mathrm{H}},& Q, & Q \\
\end{array}\right.}\right)\\
&+\frac{1}{2}\mathcal{OR}{\scriptsize\left|\begin{array}{cccc}
\mathrm{A}_{\mathrm{L}}, &  1, & N_s-Q \\
\mathrm{A}_{\mathrm{H}},& Q, & Q \\
\end{array}\right.}\times\mathcal{OR}{\scriptsize\left|\begin{array}{ccc}
0, & 1, & \lfloor KL_s\rceil-K\\
\mathrm{A}_{\mathrm{L}},&K-1, &K-1 \\
\end{array}\right.}
\end{split}
\end{equation}

\begin{equation}\label{Pe3b}
\begin{split}
\overline{P_{e_3}^\mathcal{U}}
\approx&\frac{1}{2}-\frac{1}{2}F_U(v_3)^{N_s-Q}F_U(v_4)^{\lfloor KL_s\rceil-K}-\frac{1}{3}(N_s-Q)F_U(v_3)^{N_s-Q-1}[1-F_U(v_3)]F_U(v_4)^{\lfloor KL_s\rceil-K}\\
&-\frac{1}{3}(\lfloor KL_s\rceil-K)F_U(v_4)^{\lfloor KL_s\rceil-K-1}[1-F_U(v_4)]F_U(v_3)^{N_s-Q}
\end{split}
\end{equation}
\hrule
\end{figure*}

\subsection{Approximate BER Upper Bounds of BDPIM-OSD}
In use of BDPIM-OSD, the error probability in the first OSD phase of detecting $\mathrm{A}_{\mathrm{H}}$ can be expressed as 
\setcounter{equation}{40}
\begin{equation}\label{PLH}
P_{\mathrm{LH}}=\mathcal{OR}{\scriptsize\left|\begin{array}{cccc}
\mathrm{A}_{\mathrm{L}}, &  1, & N_s-Q \\
\mathrm{A}_{\mathrm{H}},& Q, & Q \\
\end{array}\right.},
\end{equation}
and that in the second OSD phase of detecting $\mathrm{A}_{\mathrm{L}}$ is
\begin{equation}\label{P0L}
P_{0\mathrm{L}}=\mathcal{OR}{\scriptsize\left|\begin{array}{ccc}
0, & 1, & \lfloor KL_s\rceil-K\\
\mathrm{A}_{\mathrm{L}},&K-1, &K-1 \\
\end{array}\right.}.
\end{equation}
Based on these, we have the following theorem:

\begin{theorem}
The BER of BDPIM-OSD is 
approximately upper bounded by (\ref{Pe3a}) when $L$ is large.
\end{theorem}
\begin{IEEEproof}
See Appendix C.
\end{IEEEproof}

\emph{Remark and Observation:} It is observed from (\ref{Pe3a}) that $ P_{e_3}^\mathcal{U}$ is affected by various system parameters such as $\mathrm{A}_\mathrm{L}$, $\mathrm{A}_\mathrm{H}$, $K$ as well as the channel condition $h$ and $\sigma_n$. To simplify its expression, based on the approximation of the $\mathcal{OR}$ function used in Section III-B, we obtain a tractable approximate BER upper bound as (\ref{Pe3b}), where $v_3=\alpha\mathbb{E}(V_{Q:Q})$ and $v_4=\alpha\mathbb{E}(V_{K-1:K-1})$.

\begin{figure}[t]
  \centering
  \includegraphics[width=0.45\textwidth]{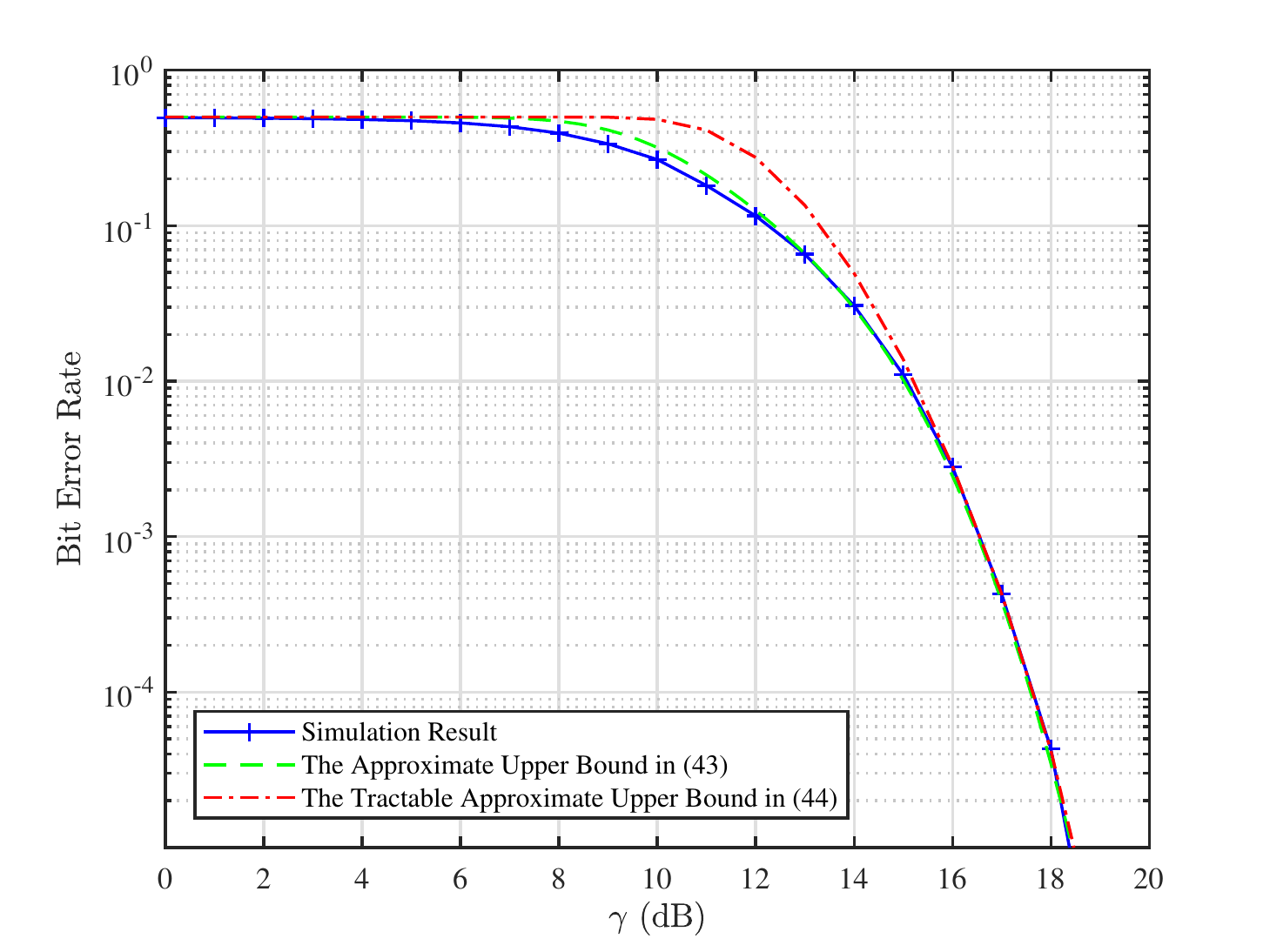}\\
 \caption{Approximate BER upper bounds of BDPIM-OSD where $h=1$, $N_s=100$, $K=10$, $\mathrm{A}_{\mathrm{H}}=2.3$, $\mathrm{A}_{\mathrm{L}}=0.86$ and $4$-BDPIM with $1$GI are employed.}
  \label{2a3}
\end{figure}

To corroborate our studies, we simulate a system using BDPIM with 1 GI and OSD. The parameters are set as: $h=1$, $N_s=100$, $K=10$, $\mathrm{A}_{\mathrm{H}}=2.3$ and $\mathrm{A}_{\mathrm{L}}=0.86$. The simulation result and the bounds are given in  Fig. \ref{2a3}. It is observed that the approximate upper bound in (\ref{Pe3a}) is tight over all SNR regimes and the tractable  approximate upper bound in (\ref{Pe3b}) is tight only in the low and high SNR regime. From these results, the accuracy in approximating the $\mathcal{OR}$ function can also be validated.

\subsection{Approximate BER Upper Bounds of BDPIM-OTD-OSD}

For BDPIM-OTD-OSD, the first phase is detecting the barrier chip of amplitude $\mathrm{A}_{\mathrm{H}}$ by using a  threshold $h\mathrm{A}_\mathrm{T}'$. Based on the similar analysis in Section III-A, $\mathrm{A}_\mathrm{T}'$ can be computed by
\setcounter{equation}{44}
\begin{equation}\label{eqAt1}
\mathrm{A}_{\mathrm{T}}'=\frac{\mathrm{A}_\mathrm{H}+\mathrm{A}_\mathrm{L}}{2}+\frac{\sigma_n^2}{h^2(\mathrm{A}_\mathrm{H}-\mathrm{A}_\mathrm{L})}\ln(K-1).
\end{equation}
Using $h\mathrm{A}_{\mathrm{T}}'$ for detection, the PEP in the first OTD phase  can be calculated by 
\begin{equation}\label{PLHN}
P_{\mathrm{LH}}'=1- (1-P_{c}')^{N_s},
\end{equation}
where $P_c'$ writes 
\begin{equation}\label{Ps}
\begin{split}
P_{c}'=&\frac{1}{K} \mathcal{Q}\left(\frac{h(\mathrm{A}_{\mathrm{H}}-\mathrm{A}_{\mathrm{L}})}{2\sigma_n}+\frac{\sigma_n}{h(\mathrm{A}_{\mathrm{H}}-\mathrm{A}_{\mathrm{L}})}\ln(K-1)\right)
\\&+\frac{K-1}{K} \mathcal{Q}\left(\frac{h(\mathrm{A}_{\mathrm{H}}-\mathrm{A}_{\mathrm{L}})}{2\sigma_n}-\frac{\sigma_n}{h(\mathrm{A}_{\mathrm{H}}-\mathrm{A}_{\mathrm{L}})}\ln(K-1)\right).
\end{split}
\end{equation}
The error detection probability in the second OSD phase is the same as that in (\ref{P0L}). Based on these, we have the following theorem:
\begin{theorem}
The BER of BDPIM-OTD-OSD is approximately upper bounded by (\ref{Pe4a}) when $L$ is large.
\end{theorem}
\begin{IEEEproof}
See Appendix D.
\end{IEEEproof}
\begin{figure*}
\setcounter{equation}{47}
\begin{equation}\label{Pe4a}
\begin{split}
 P_{e_4}^\mathcal{U}
\approx&\frac{2-2(1-P_c')^{N_s}-LP_c'(1-P_c')^{N_s-1}}{4}\left(1-\mathcal{OR}{\scriptsize\left|\begin{array}{ccc}
0, & 1, & \lfloor KL_s\rceil-K\\
\mathrm{A}_{\mathrm{L}},&K-1, &K-1 \\
\end{array}\right.}\right)\\
&+\frac{1}{6}\left(\mathcal{OR}{\scriptsize\left|\begin{array}{ccc}
0, & 1, & \lfloor KL_s\rceil-K\\
\mathrm{A}_{\mathrm{L}},&K-1, &K-1 \\
\end{array}\right.}+2\mathcal{OR}{\scriptsize\left|\begin{array}{ccc}
0, & 2, & \lfloor KL_s\rceil-K\\
\mathrm{A}_{\mathrm{L}},&K-2, &K-1 \\
\end{array}\right.}\right)(1-P_c')^{N_s}\\
&+\frac{1}{2}\left[1-(1-P_c')^{N_s}\right]\mathcal{OR}{\scriptsize\left|\begin{array}{ccc}
0, & 1, & \lfloor KL_s\rceil-K\\
\mathrm{A}_{\mathrm{L}},&K-1, &K-1 \\
\end{array}\right.}
\end{split}
\end{equation}

\begin{equation}\label{Pe4b}
\begin{split}
\overline{P_{e_4}^\mathcal{U}}
\approx&\frac{1}{2}-\frac{1}{2}(1-P_c')^{N_s}F_U(v_4)-\frac{1}{4}LP_c'(1-P_c')^{N_s-1}F_U(v_4)^{\lfloor KL_s\rceil-K}\\
&-\frac{1}{3}(\lfloor KL_s\rceil-K)F_U(v_4)^{\lfloor KL_s\rceil-K-1}[1-F_U(v_4)](1-P_c')^{N_s}
\end{split}
\end{equation}
\hrule
\end{figure*}

\begin{figure}[t]
  \centering
  \includegraphics[width=0.45\textwidth]{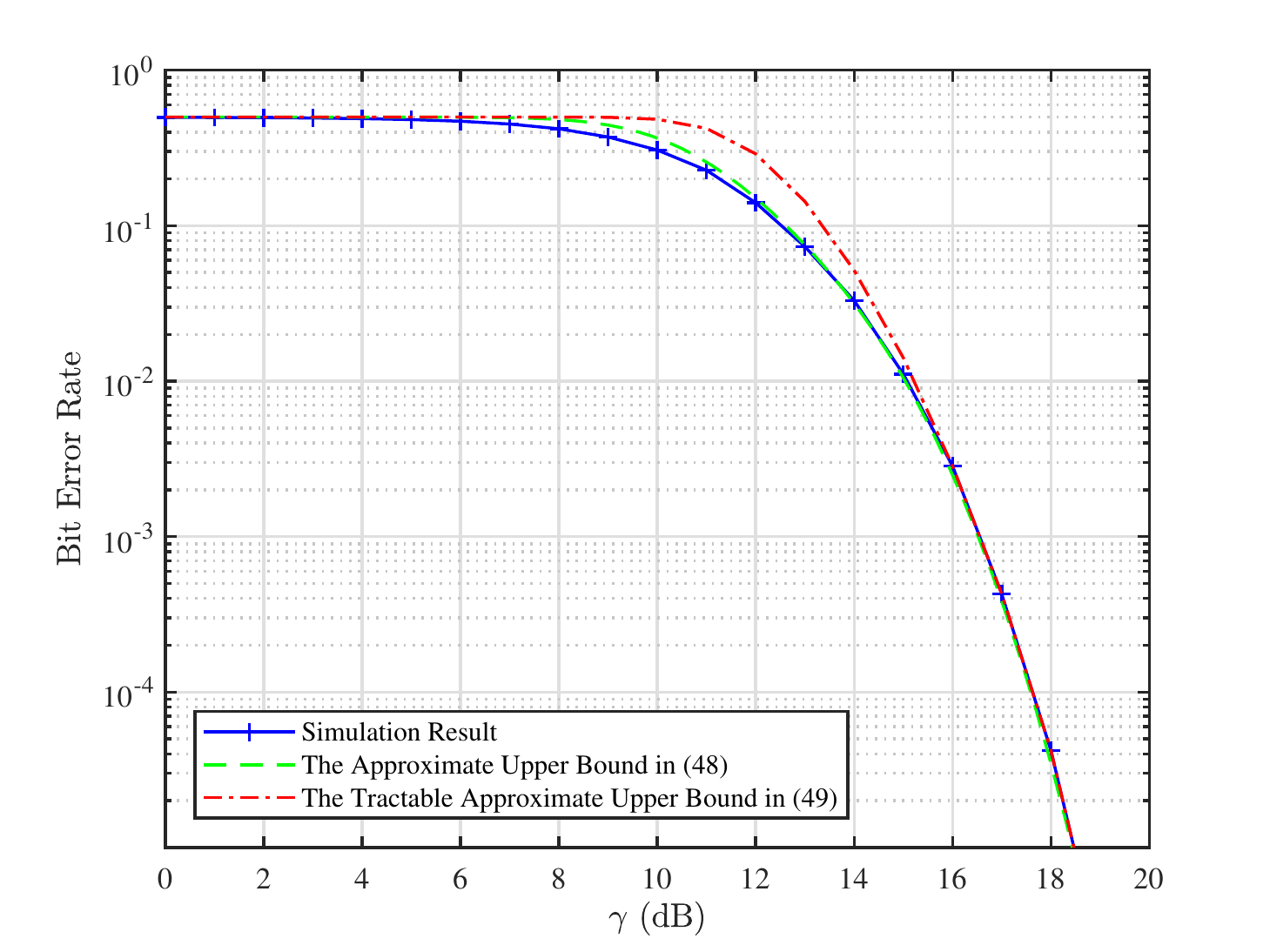}\\
 \caption{Approximate BER upper bounds of BDPIM-OTD-OSD where $h=1$, $N_s=100$, $K=10$, $\mathrm{A}_{\mathrm{H}}=2.3$, $\mathrm{A}_{\mathrm{L}}=0.86$ and $4$-BDPIM with $1$GI are employed.}
  \label{2a4}
\end{figure}

\emph{Remark and Observation:}
Similarly, using the approximation of the $\mathcal{OR}$ function used in Section III-B, we obtain a tractable approximate BER upper bound as (\ref{Pe4b}).
 Under the same simulation setups as that in Section III-C, we simulate BDPIM-OTD-OSD. The simulation result and the bounds are given in  Fig. \ref{2a4}. Similarly, it is also observed that the upper bound in (\ref{Pe4a}) are quite tight over all SNR regimes and that in (\ref{Pe4a}) is close to the BER performance in the low and high SNR regime.

\section{Simulation and Discussion}
This section presents the simulation results and has four subsections. In the first subsection, we show the superiority of the proposed OSD over OTD and MLSD.  In the second subsection,  we show the superiority of BDPIM in uncoded and coded systems. In the third subsection, we study the impact of the parameters on the performance of the proposed BDPIM. In the last subsection, we investigate the extension to Gamma-Gamma turbulence channels.

\subsection{Superiority of OSD}
\begin{figure}[t]
  \centering
  \includegraphics[width=0.45\textwidth]{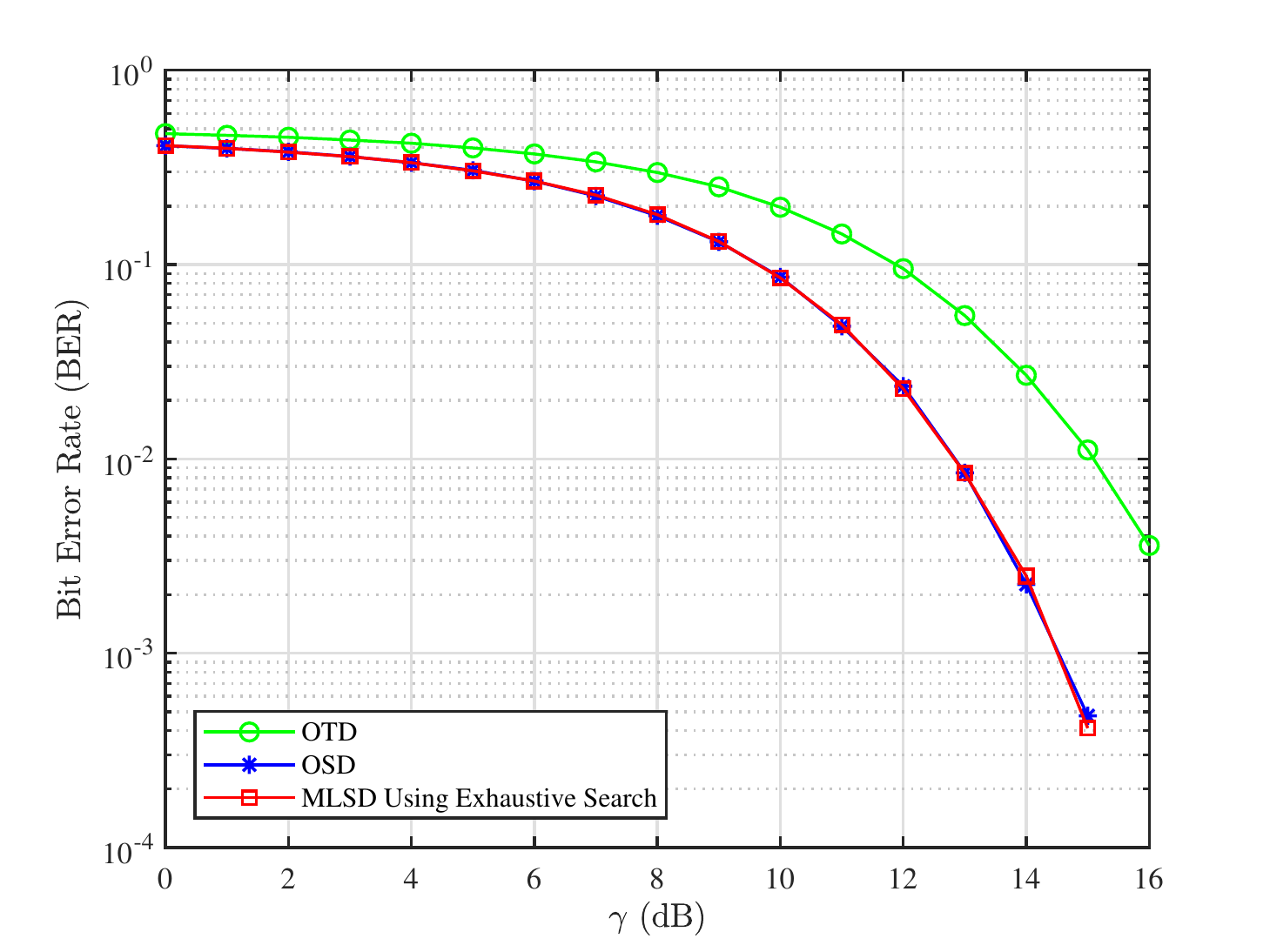}\\
 \caption{Comparisons among OTD, OSD and MLSD with exhaustive search in terms of BER  where $N_s=4$ and $4$-DPIM with $1$ GI are employed.}
  \label{MLSDa}
\end{figure}

\begin{figure}[t]
  \centering
  \includegraphics[width=0.45\textwidth]{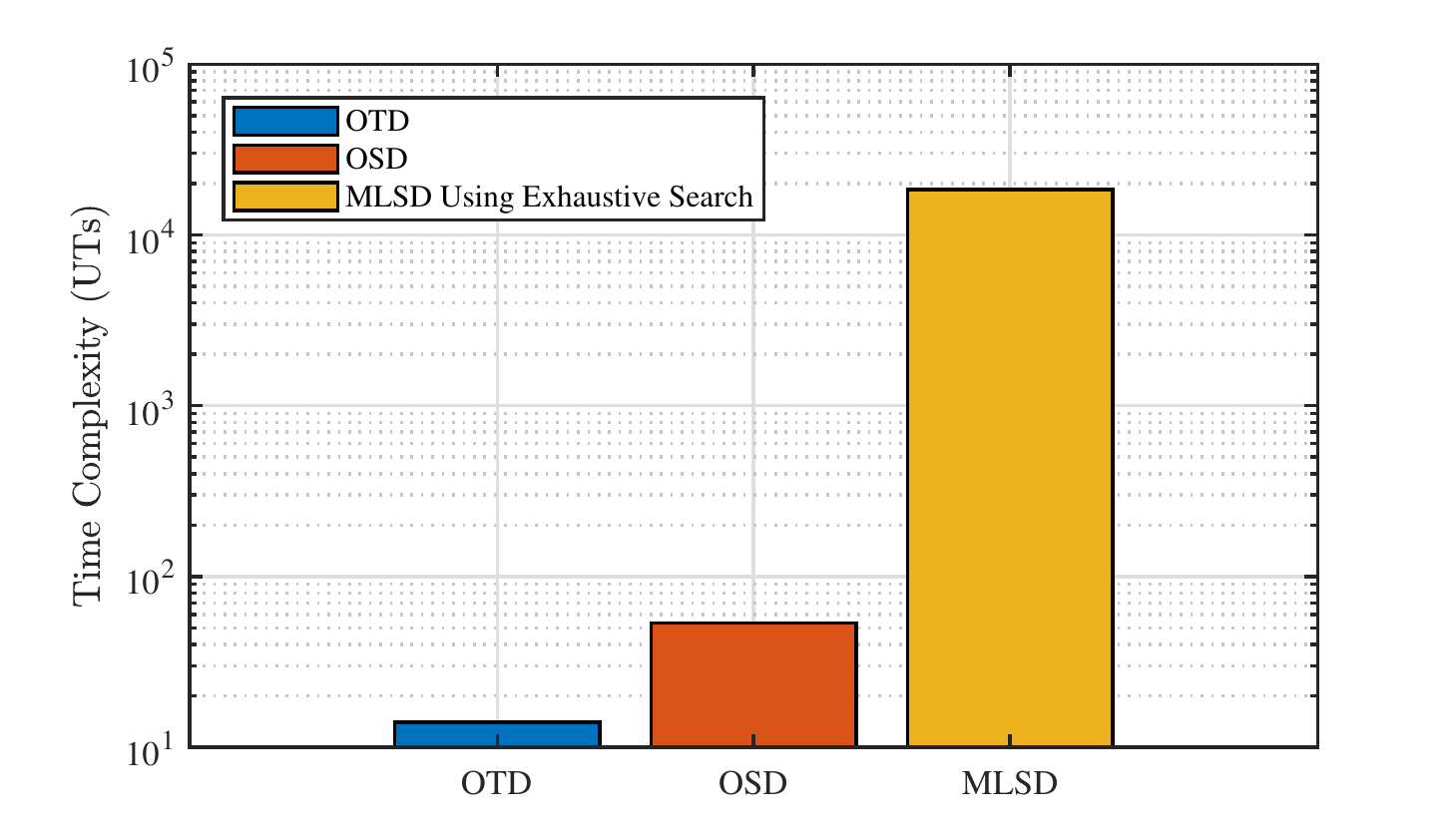}\\
 \caption{Comparisons among OTD, OSD and MLSD in terms of time complexity, where a UT is the amount of time it takes for a multiplication.}
  \label{MLSDb}
\end{figure}

Firstly, we compare the proposed OSD with OTD and MLSD using exhaustive search in terms of BER and time complexity. Because MLSD using exhaustive search has prohibitive high complexity when the sequence is long, we compare these schemes in a system transmitting small packets containing $4$ symbols per packet. In the comparison, $h=1$ and $4$-DPIM with $1$ GI are employed. We run $10^5$ Monte Carlo simulations and evaluate the BER performance as well as the average time complexity in UTs. The BER performance comparison is shown in Fig. \ref{MLSDa}, from which we observe that OSD exhibits the same performance as MLSD using exhaustive search and considerably outperforms OTD over a wide SNR regime. The average time complexity  is included in Fig. \ref{MLSDb}, which shows that OSD is much more computationally efficient than MLSD using exhaustive search.


\subsection{Superiority of BDPIM}
 \begin{figure}[t]
  \centering
  \includegraphics[width=0.45\textwidth]{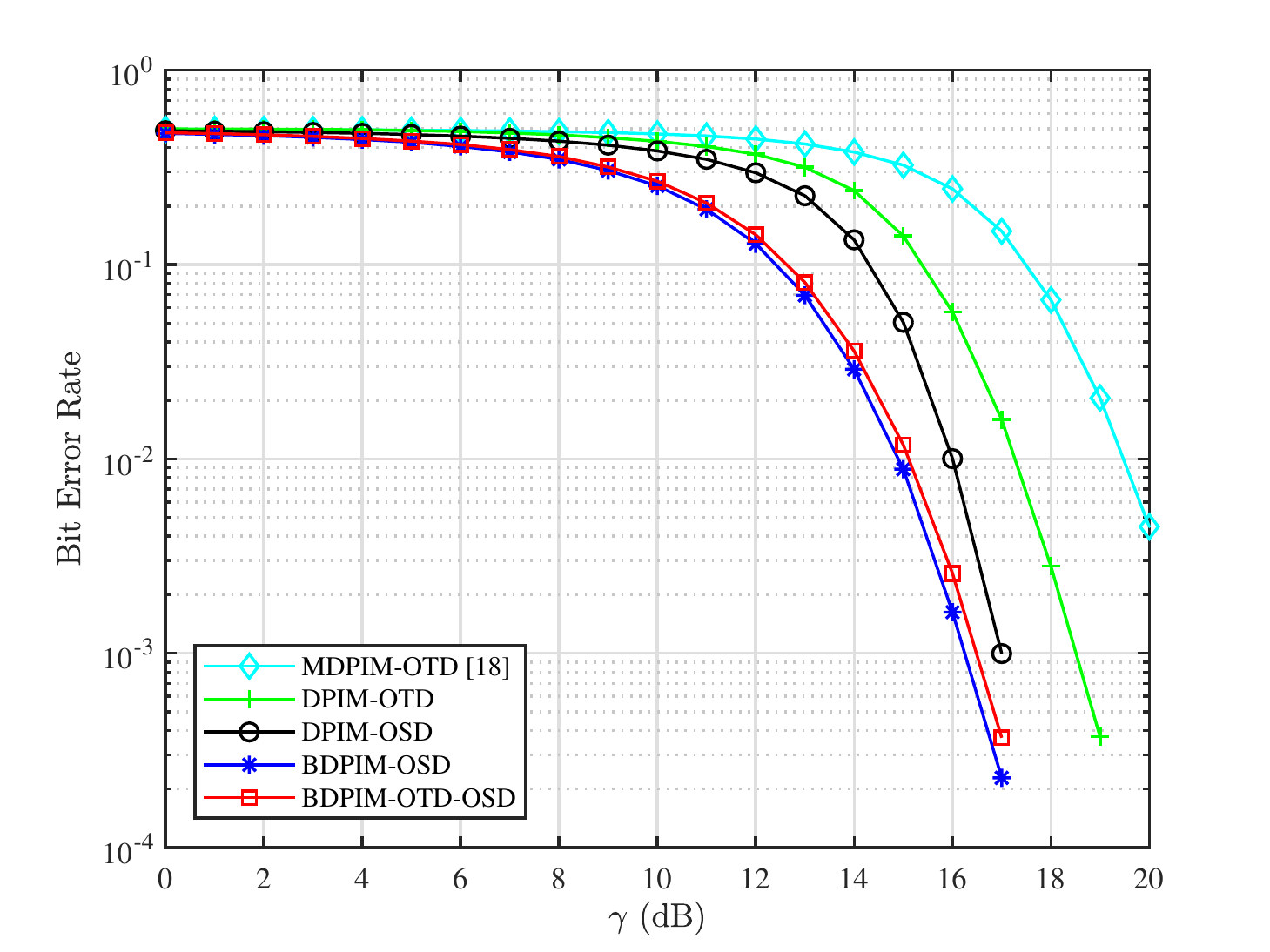}\\
 \caption{Comparisons among MDPIM-OTD \cite{Ghassemlooy2006}, DPIM-OTD, DPIM-OSD, BDPIM-OSD and BDPIM-OTD-OSD in terms of uncoded BER performance.}
  \label{BDPIM}
\end{figure}

Secondly, we compare BDPIM with DPIM and MDPIM \cite{Ghassemlooy2006} using various detectors in uncoded and coded systems. The parameters $h=1$, $N_s=100$, $K=10$, and $4$-ary BDPIM with $1$ GI are employed. In coded systems, an interleaver of length $K\log M=20$ is used. To show the proposed BDPIM in association with high-rate FEC codes, a rate-$1/2$ convolution code is adopted and its code generator can be expressed as $[g_0,g_1]^T=[1+d+d^2,1+d^2]$. Additionally, the Viterbi hard-decision decoder is used for decoding. For $\mathrm{A}_{\mathrm{H}}$ and $\mathrm{A}_{\mathrm{L}}$, we adopt the optimal solution obtained in Section IV-C. The uncoded simulation results are included Fig. \ref{BDPIM}. It demonstrates that BDPIM-OSD exhibits the best BER performance in all depicted SNR regimes. It slightly outperforms BDPIM-OTD-OSD, which is applicable in large packet cases. Compared with DPIM-OSD, BDPIM-OSD offers a significant gain at the medium SNR regime. The gain shrinks as SNR goes larger. The reason is as follows. The BER performance is affected by two factors, the PER and the expected BER (affected by error propagation) in an erroneous packet. BDPIM uses some power to offer built-in barriers, which slightly increases PER and greatly reduces the expected BER in an erroneous packet. At the medium SNR regime, the second factor affects the system performance more. As SNR increases, the impacts of two factors on BER performance become comparable. Therefore, the gain shrinks as SNR increases. From Fig. \ref{BDPIM}, we also observe that BDPIM-OSD greatly outperforms DPIM-OTD by $2$ dB at a BER of $10^{-2}$. Additionally, we compare BDPIM with the MDPIM in \cite{Ghassemlooy2006}, which uses multiple amplitudes to additionally transmit information. Note that as the average symbol duration of MDPIM is shorter than BDPIM (shown in Table II), we simulate MDPIM under the same average power per chip for fairness. It is observed that BDPIM-OSD considerably outperforms MDPIM by more than $4$ dB. This is reasonable because MDPIM targets to improve the transmission rate while BDPIM aims to enhance the system reliability.

 \begin{figure}[t]
  \centering
  \includegraphics[width=0.45\textwidth]{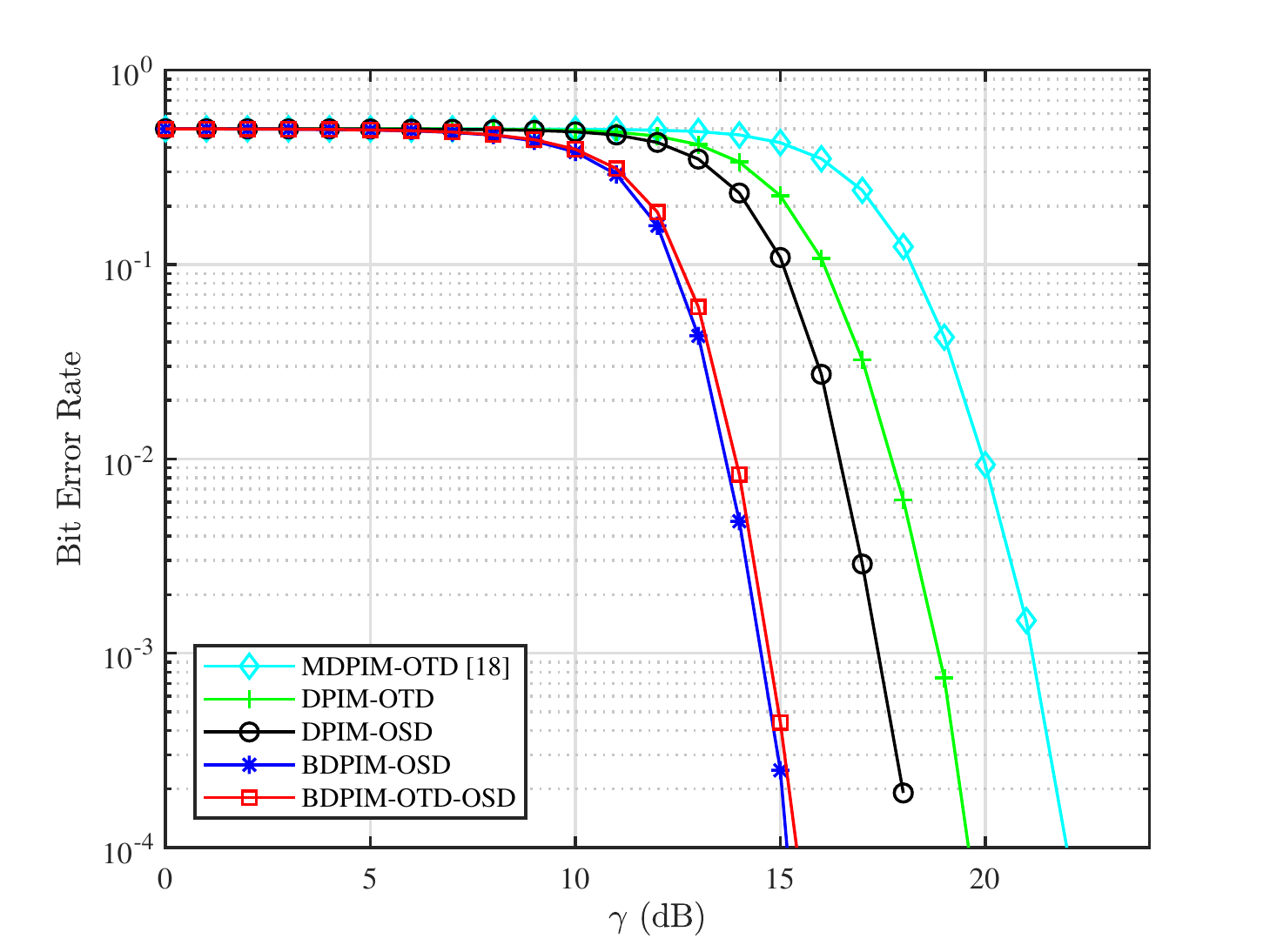}\\
 \caption{Comparisons among MDPIM-OTD \cite{Ghassemlooy2006}, DPIM-OTD, DPIM-OSD, BDPIM-OSD and BDPIM-OTD-OSD in terms of coded BER performance.}
  \label{codedBDPIM}
\end{figure}

The comparison made in coded systems is demonstrated in Fig. \ref{codedBDPIM}.  It shows that BDPIM-OSD still achieves the best performance and is slightly better than BDPIM-OTD-OSD. Both of them greatly outperform the others. Note that the gain brought by BDPIM-OSD does not shrink as SNR increases in coded systems compared with DPIM-OSD. On the contrary, the gain goes larger as SNR increases. This is because the length of block error is short in BDPIM by limiting the error propagation which enables the functionality of the rate-1/2 FEC code in improving BER performance. For DPIM, even though its PER is low in the high SNR regime (shown in uncoded systems), the error bits in an erroneous packet are too many to be corrected. As a result, the gain becomes more significant. Moreover, Fig. \ref{codedBDPIM} also demonstrates that the gains brought by BDPIM-OSD  are also more considerable than that in Fig. \ref{BDPIM} compared with DPIM-OTD and MDPIM-OTD.
Therefore, we conclude that the gain brought by BDPIM is more considerable in coded systems. 

\subsection{Parameter Impact on the Performance of BDPIM}
Given the number of symbols contained in a packet $N_s$, the parameters of BDPIM include $K$, $\mathrm{A}_{\mathrm{L}}$ and $\mathrm{A}_{\mathrm{H}}$.  The parameter optimization problems of BDPIM with different detectors can be expressed as
\begin{equation}\label{Problem}
\begin{split}
K^*,\mathrm{A}_{\mathrm{L}}^*, \mathrm{A}_{\mathrm{H}}^*=&\arg\min_{K,\mathrm{A}_{\mathrm{L}},\mathrm{A}_{\mathrm{H}}} P_{e_i}^{\mathcal{U}}(K,\mathrm{A}_{\mathrm{L}},\mathrm{A}_{\mathrm{H}}), ~i=3,4\\
\mathrm{subject~to:~}& 0<\mathrm{A}_{\mathrm{L}}<\mathrm{A}_{\mathrm{H}}\textrm{~and~} (\ref{Con1}).\\
\end{split}
\end{equation}
Given $K$ and $\mathrm{A}$, the constraint in (\ref{Con1}) can be expressed as 
\begin{equation}
\mathrm{A}_{\mathrm{H}}=K\mathrm{A}-(K-1)\mathrm{A}_{\mathrm{L}}.
\end{equation}
Thus, the problem in (\ref{Problem}) becomes a one-dimensional search problem of finding the optimal $\mathrm{A}_{\mathrm{L}}$ from $(0,\mathrm{A})$. As the first derivative of $P_{e_i}^{\mathcal{U}}(K,\mathrm{A}_{\mathrm{L}},\mathrm{A}_{\mathrm{H}}), ~i=3,4$ with respect to $\mathrm{A}_{\mathrm{L}}$ has no simple expression, it is difficult to derive the closed-form solution. Therefore, we resort to using numerical search methods such as the bisection search algorithm \cite{Boyd2004} in this paper. 

 \begin{figure}[t]
  \centering
  \includegraphics[width=0.45\textwidth]{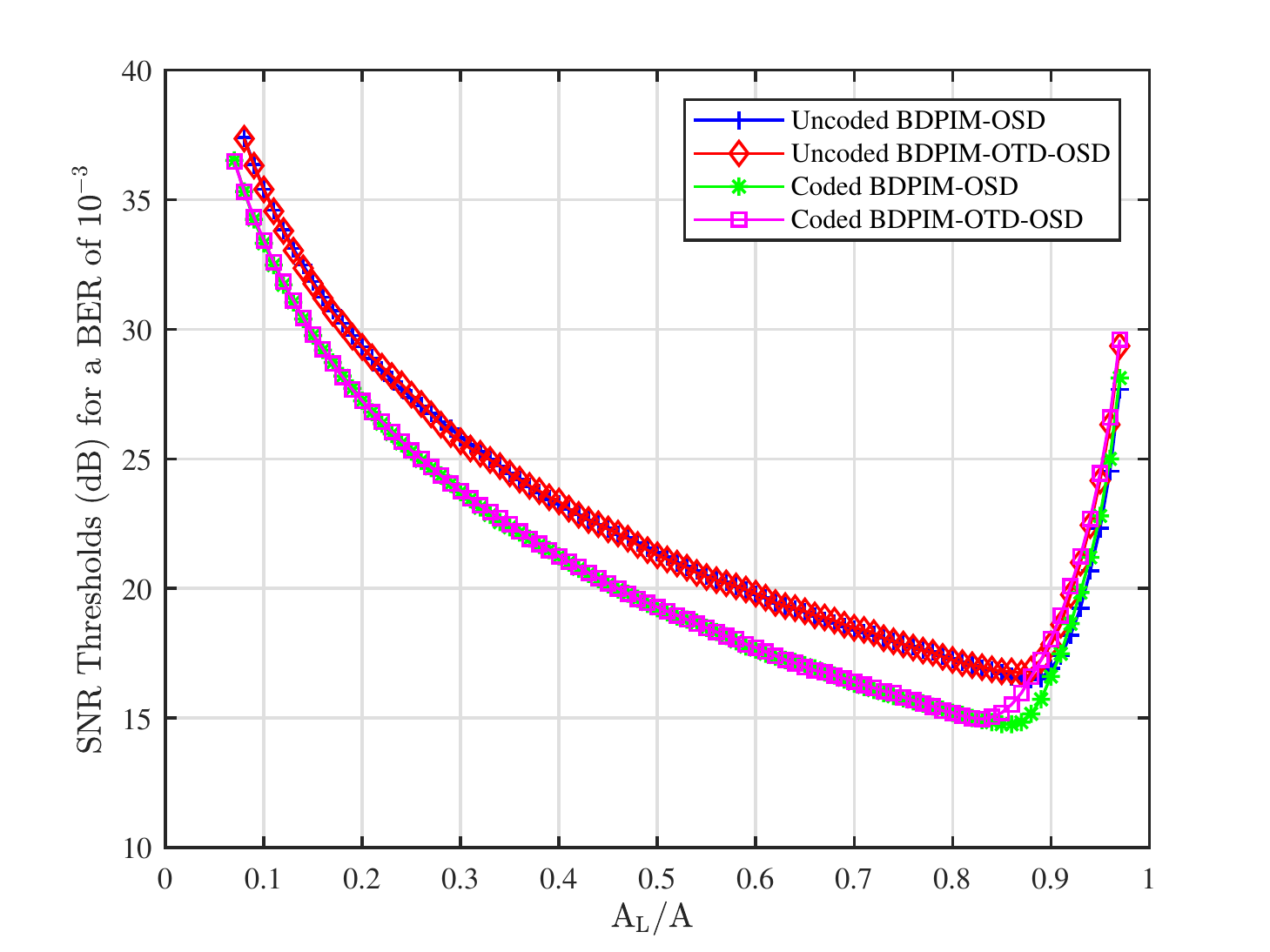}\\
 \caption{SNR thresholds for a BER of $10^{-3}$ versus different $\mathrm{A}_{\mathrm{L}}/\mathrm{A}$.}
  \label{3a_new}
\end{figure}

\begin{figure}[t]
  \centering
  \includegraphics[width=0.45\textwidth]{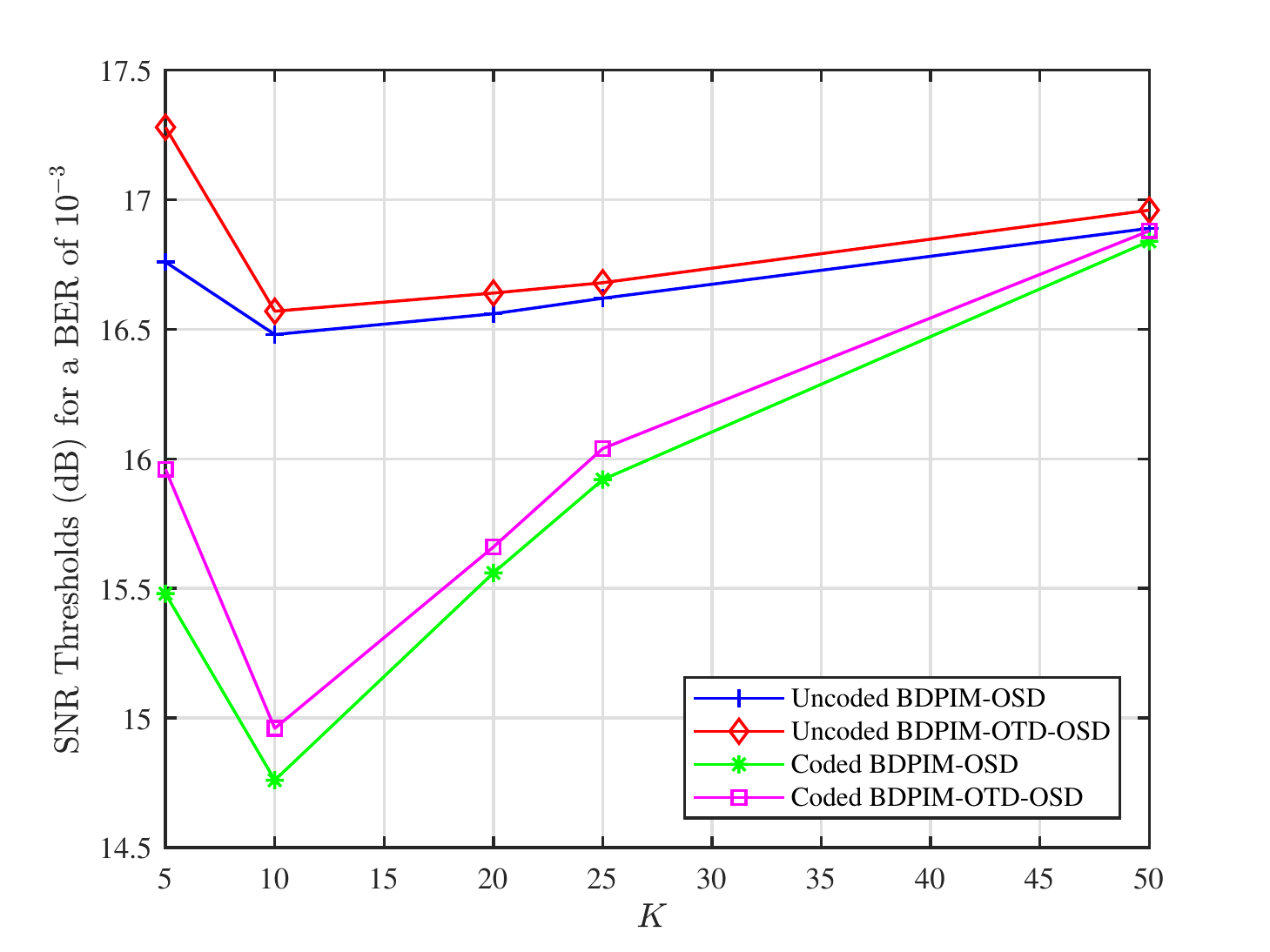}\\
 \caption{SNR thresholds for a BER of $10^{-3}$ versus different $K$.}
  \label{3b_new}
\end{figure}

To show how these parameters affect the performance of BDPIM, we simulate the system with a varying parameter. By setting $\mathrm{A}_{\mathrm{L}}/\mathrm{A}$ to vary from $0$ to $1$ by a step of $0.01$, $h=1$, $N_s=100$ and $K=10$, we obtain the SNR thresholds for a BER of $10^{-3}$ in coded and uncoded BDPIM systems as shown in Fig. \ref{3a_new}.   It shows that the optimal value of $\mathrm{A}_{\mathrm{L}}$ in coded systems is smaller than that in uncoded systems. That is, the optimal value of $\mathrm{A}_{\mathrm{H}}$ in coded systems is larger than that in uncoded systems. It can be explained as follows. The FEC code can correct the errors of the second phase conditioned the first phase is correctly detected. Therefore, the larger $\mathrm{A}_{\mathrm{H}}$ should be adopted in coded systems to ensure the correctness of the first detection phase.
Additionally, by setting $K$ to vary in $\{5,10,20,25,50\}$, $h=1$, $N_s=100$ and $\mathrm{A}_{\mathrm{L}}$, $\mathrm{A}_{\mathrm{H}}$ as the searched optimal values, we obtain the SNR thresholds for a BER of $10^{-3}$ in coded and uncoded BDPIM systems as shown in Fig. \ref{3b_new}. It shows that the values of $K$ affect coded systems more than uncoded systems and the optimal ones for all systems are $10$. Specifically, the SNR thresholds vary from $16.5$ dB to $17.3$ dB in uncoded systems as $K$ varies, while the SNR thresholds vary in a much larger range from $14.8$ dB to $16.8$ dB in coded systems as $K$ varies. This is because $K$ affects the length of block error and determines whether the block error can be corrected or not.
\subsection{Extension to Gamma-Gamma Turbulence Channels}
In this part, we consider the extension of the proposed designs to other channels. The Gamma-Gamma turbulence channels are taken as an example. The ergodic  approximate  upper bound for any given  channel coefficient distribution can be expressed as
\begin{equation}\label{EPe}
\widetilde{P_{e_{i}}^{\mathcal{U}}}=\int_{0}^{+\infty}P_{e_{i}}^{\mathcal{U}}(h)f_{h}{(h)} dh, ~i=1,2,3,4.
\end{equation}
For Gamma-Gamma turbulence channels, the channel coefficient distribution can be expressed as\cite{Jaiswal2017}
\begin{equation}\label{pdfH}
f_{h}{(h)}=\frac{2(\lambda\mu)^{\frac{\lambda+\mu}{2}}}{\Gamma(\lambda)\Gamma(\mu)}h^{\frac{\lambda+\mu}{2}}K_{\lambda-\mu}\left(\sqrt{\lambda\mu h}\right)
\end{equation}
where $\Gamma(\cdot)$ and $K_{\lambda-\mu}(\cdot)$ are the Gamma function and the modified Bessel function, respectively. In (\ref{pdfH}),  $\lambda$ and $\mu$ are the turbulence parameters that characterize the irradiance fluctuations.
 By inserting (\ref{pdfH}) into (\ref{EPe}), one can readily extend to the analytical results given in Theorems 1-4 to Gamma-Gamma turbulence channels. 

\begin{figure}[t]
  \centering
  \includegraphics[width=0.45\textwidth]{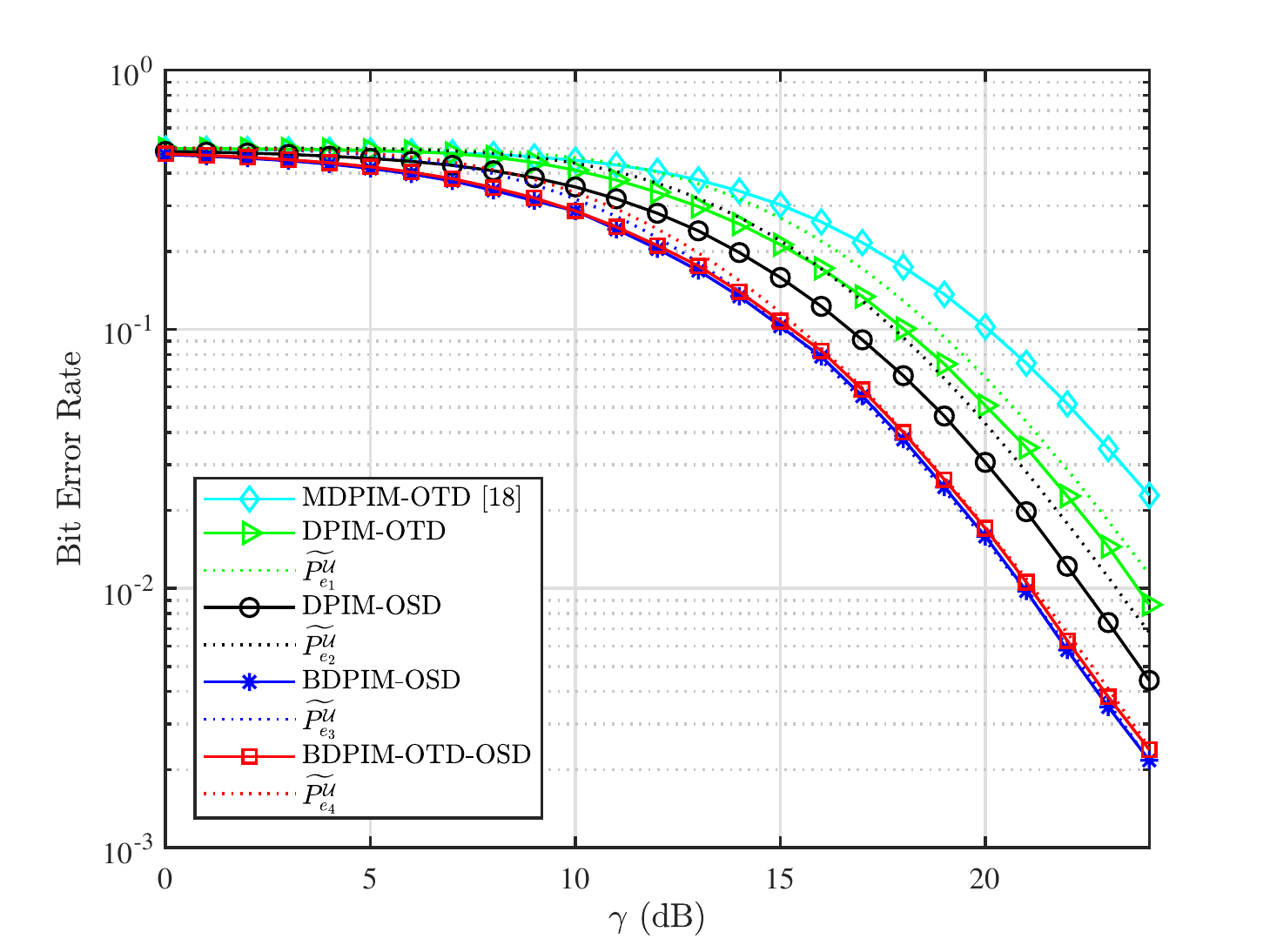}\\
 \caption{Comparisons among MDPIM-OTD \cite{Ghassemlooy2006}, DPIM-OTD, DPIM-OSD, BDPIM-OSD and BDPIM-OTD-OSD in terms of BER performance over Gamma-Gamma weak turbulence channel.}
  \label{GGBDPIM}
\end{figure}

We simulate the system under weak turbulence condition, where $\lambda$ and $\mu$ are set as $11.6$ and $10.1$ \cite{Jaiswal2017}, respectively. The  results are included in Fig. \ref{GGBDPIM}. Results show that the proposed BDPIM and OSD can bring about $4$-dB gain compared with the widely adopted DPIM-OTD over a wide SNR regime.

\section{Conclusions}
This paper proposed an OSD and a BDPIM to improve the system reliability of DPIM. With  low complexity, OSD achieves the same system performance as the optimal MLSD in terms of BER. Through specific power allocation, BDPIM makes the error propagation bounded and enables the association with high-rate FEC codes, which greatly improves the system reliability. The  approximate BER upper bounds of four schemes including DPIM-OTD, DPIM-OSD, BDPIM-OSD and BDPIM-OTD-OSD were derived. The approximate upper bounds were shown tight over all SNR regimes, but they are expressed with complicated integrals. To fix this, this paper also provided tractable approximate upper bounds, which are expressed in closed-form expressions. We studied how parameter settings affect the performance by simulations.  We compared all schemes over AWGN channel and over Gamma-Gamma turbulence channels. Result showed that the proposed BDPIM and OSD can bring considerable performance gain and enhance the system reliability significantly.

\appendices

\section{Proof of Theorem 1}
For DPIM-OTD, the expected BER in an erroneous packet  can be expressed as
\begin{equation}
\overline{\mathsf{BER}}_1=\beta_{1}\overline{\mathsf{BER}}_{1,1}+(1-\beta_{1})\overline{\mathsf{BER}}_{1,o}.
\end{equation}
where $\beta_1$ is the probability of there being a chip error conditioned a packet is wrongly detected, $\overline{\mathsf{BER}}_{1,1}$ represents the expected BER in an erroneous packet with a single-chip error and $\overline{\mathsf{BER}}_{1,o}$ stands for the expected BER in an erroneous packet with more than one chip error.
For either a \emph{false alarm error} or an \emph{erasure error} in OTD, the error not only corrupts the bits directly associated with that chip, but also shifts the bits that follow those bits, such that
\begin{equation}\label{BER1}
\overline{\mathsf{BER}}_{1,1}=\sum_{t=1}^{L}\frac{N_s-\mathcal{N}(t)}{2LN_s},
\end{equation}
where $t$ is the error position and $\mathcal{N}(t)$ stands for the number of symbols (i.e., the number of $\mathrm{A}$s)  between the $1$st and the $t$th chip. Since the average symbol duration is $L_s$, $\mathcal{N}(t)\approx t/L_s$ and $L\approx N_sL_s$. Substituting these into (\ref{BER1}), we can get $\overline{\mathsf{BER}}_{1,1}\approx (L-1)/4L\approx1/4$. 
For the other cases,  $\overline{\mathsf{BER}}_{1,o}$ is less than $1/2$ (the BER of random guess).
As to the probability $\beta_1$, it can be calculated by
\begin{equation}\label{beta1}
\begin{split}
\beta_1=\frac{LP_c(1-P_c)^{L-1}}{1-(1-P_c)^L}.
\end{split}
\end{equation}
Summarizing above, $\overline{\mathsf{BER}}_1\leq \frac{2-\beta_1}{4}$ and $P_{e_1}\leq \frac{2-\beta_1}{4}P_{p_1}$. By substituting $\beta_1$ in (\ref{beta1}) and $P_{p_1}$ in (\ref{Pp1}), Theorem 1 is proved.

\section{Proof of Theorem 2}
As stated in Section III-B, the chip errors of OSD occur in pairs.
For DPIM-OSD, the expected BER in an erroneous packet  can be expressed as
\begin{equation}
\overline{\mathsf{BER}}_2=\beta_{2}\overline{\mathsf{BER}}_{2,1}+(1-\beta_{2})\overline{\mathsf{BER}}_{2,o}.
\end{equation}
where $\beta_2$ represents the probability of there being a pair of chip errors conditioned a packet is wrongly detected, $\overline{\mathsf{BER}}_{2,1}$ denotes the expected BER in an erroneous packet with a pair of chip errors and $\overline{\mathsf{BER}}_{2,o}$ is the expected BER in an erroneous packet with more than one pair of chip errors.
When a single pair of chip errors occur in OSD, the errors will corrupt the bits between these two chips and $\overline{\mathsf{BER}}_{2,1}$ can be expressed as
\begin{equation}
\overline{\mathsf{BER}}_{2,1}=\sum_{t_1=1}^{L}\sum_{t_2=t_1+1}^{L}\frac{\mathcal{N}(t_2)-\mathcal{N}(t_1)}{L(L-1)N_s},
\end{equation}
where $t_1$ and $t_2$ are the error positions. Based on the approximations $\mathcal{N}(t_2)\approx t_2/L_s$, $\mathcal{N}(t_1)\approx t_1/L_s$ and $L\approx N_sL_s$, the BER of DPIM-OSD can be obtained as
\begin{equation}
\overline{\mathsf{BER}}_{2,1}\approx \frac{2L^2-7L+5}{12L^2}\approx \frac{1}{6}.
\end{equation}
This indicates that averagely, one sixth of bits are in error in an erroneous packet for this case. 
For the other cases, $\overline{\mathsf{BER}}_{2,o}$ is less than $1/2$ (the BER of random guess), i.e., $\overline{\mathsf{BER}}_{2,o}\leq \frac{1}{2}$.
As to the probability $\beta_2$, it can be expressed as
\begin{equation}\label{beta2}
\begin{split}
\beta_2&=\frac{P_{p_2}-\mathrm{Pr}\{U_{2:L-N_s}>V_{N_s-1:N_s}\}}{P_{p_2}}\\
&\approx\frac{\mathcal{OR}{\scriptsize\left|\begin{array}{cccc}
0, & 1, & L-N_s \\
\mathrm{A},&N_s, & N_s \\
\end{array}\right.}-\mathcal{OR}{\scriptsize\left|\begin{array}{cccc}
0, & 2, & L-N_s \\
\mathrm{A},&N_s-1, & N_s \\
\end{array}\right.}}{\mathcal{OR}{\scriptsize\left|\begin{array}{cccc}
0, & 1, & L-N_s \\
\mathrm{A},&N_s, & N_s \\
\end{array}\right.}},
\end{split}
\end{equation}
because $P_{p_2}$ in (\ref{eqPp2}) is the probability that there are at least one pair of chip errors in an erroneous packet, $\mathrm{Pr}\{U_{2:L-N_s}>V_{N_s-1:N_s}\}$ represents the probability that there are at least two pairs of chip errors in an erroneous packet and the difference between the two terms is the probability of a single pair of chip errors.
Summarizing above, $\overline{\mathsf{BER}}_2\leq \frac{(3-2\beta_2)}{6}$ and $P_{e_2}\leq \frac{(3-2\beta_2)}{6}P_{p_2}$. By substituting $\beta_2$ in (\ref{beta2}) and $P_{p_2}$ in (\ref{Pp2}), Theorem 2 is proved.

\section{Proof of Theorem 3}
There are 5 events for all error cases of BDPIM-OSD.

Event 1: The first OSD phase is wrongly detected, the second OSD phase is correctly detected and there are only a single pair of chip errors in the first OSD phase. The probability  of this event is $\beta_{3,a}P_{\mathrm{LH}}(1-P_{0\mathrm{L}})$ and the expected BER in a packet of this event is denoted by  $\overline{\mathsf{BER}}_{3,1}$. According to the similar analysis in Appendix B, $\overline{\mathsf{BER}}_{3,1}\approx1/6$ and $\beta_{3,a}$ can be expressed as
\begin{equation}
\begin{split}
&\beta_{3,a}=\frac{\mathcal{OR}{\scriptsize\left|\begin{array}{cccc}
\mathrm{A}_{\mathrm{L}}, & 1, & N_s-Q \\
\mathrm{A}_{\mathrm{H}},&Q, & Q \\
\end{array}\right.}-\mathcal{OR}{\scriptsize\left|\begin{array}{cccc}
\mathrm{A}_{\mathrm{L}}, & 2, & N_s-Q \\
\mathrm{A}_{\mathrm{H}},&Q-1, & Q \\
\end{array}\right.}}{\mathcal{OR}{\scriptsize\left|\begin{array}{cccc}
\mathrm{A}_{\mathrm{L}}, & 1, & N_s-Q \\
\mathrm{A}_{\mathrm{H}},&Q, & Q \\
\end{array}\right.}}.\\
\end{split}
\end{equation}

Event 2: The first OSD phase is correctly detected, the second OSD phase is wrongly detected and there are only a single pair of of chip errors in the second OSD phase. The probability of this event is $\beta_{3,b}(1-P_{\mathrm{LH}})P_{0\mathrm{L}}$ and the expected BER in a packet of this event is denoted by  $\overline{\mathsf{BER}}_{3,2}$. According to the similar analysis in Appendix B, $\overline{\mathsf{BER}}_{3,2}\approx1/6$ and $\beta_{3,b}$ can be expressed as
\begin{equation}
\begin{split}
&\beta_{3,b}=\frac{\mathcal{OR}{\scriptsize\left|\begin{array}{ccc}
0, & 1, & \lfloor KL_s\rceil-K\\
\mathrm{A}_{\mathrm{L}},&K-1, &K-1 \\
\end{array}\right.}-\mathcal{OR}{\scriptsize\left|\begin{array}{ccc}
0, & 2, & \lfloor KL_s\rceil-K\\
\mathrm{A}_{\mathrm{L}},&K-2, &K-1 \\
\end{array}\right.}}{\mathcal{OR}{\scriptsize\left|\begin{array}{ccc}
0, & 1, & \lfloor KL_s\rceil-K\\
\mathrm{A}_{\mathrm{L}},&K-1, &K-1 \\
\end{array}\right.}},\\
\end{split}
\end{equation}

Event 3: The first OSD phase is wrongly detected, the second OSD phase is correctly detected and there are more than one pair of chip errors in the first OSD phase. The probability  of this event is $(1-\beta_{3,a})P_{\mathrm{LH}}(1-P_{0\mathrm{L}})$ and the expected BER in a packet of this event is denoted by  $\overline{\mathsf{BER}}_{3,3}$, which is less than $1/2$ (the BER of random guess). 

Event 4: The first OSD phase is correctly detected, the second OSD phase is wrongly detected and there are more than one pair of chip errors in the second OSD phase. The probability  of this event is $(1-\beta_{3,b})(1-P_{\mathrm{LH}})P_{0\mathrm{L}}$ and the expected BER in a packet of this event is denoted by  $\overline{\mathsf{BER}}_{3,4}$, which is less than $1/2$ (the BER of random guess). 

Event 5: Both OSD phases are wrongly detected. The probability  of this event is $P_{\mathrm{LH}}P_{0\mathrm{L}}$ and the expected BER in a packet of this event is denoted by  $\overline{\mathsf{BER}}_{3,5}$, which is less than $1/2$ (the BER of random guess). 

Summarizing all above and substituting $P_{\mathrm{LH}}$ in (\ref{PLH}) and, $P_{0\mathrm{L}}$ in (\ref{P0L}), Theorem 3 can be derived.

\section{Proof of Theorem 4}
There are 5 events for all error cases of BDPIM-OTD-OSD.

Event 1: The first OTD phase is wrongly detected, the second OSD phase is correctly detected and there is only a single chip error in the first OTD phase. The probability  of this event is $\beta_{4,a}P_{\mathrm{LH}}'(1-P_{0\mathrm{L}})$ and the expected BER in a packet of this event is denoted by  $\overline{\mathsf{BER}}_{4,1}$. According to the similar analysis in Appendix A, $\overline{\mathsf{BER}}_{4,1}\approx1/4$ and $\beta_{4,a}$ can be expressed as
\begin{equation}
\begin{split}
&\beta_{4,a}=\frac{N_sP_c'(1-P_c')^{N_s-1}}{1-(1-P_c')^N_s}.
\end{split}
\end{equation}

Event 2: The first OTD phase is correctly detected, the second OSD phase is wrongly detected and there are only a single pair of of chip errors in the second OSD phase. The probability of this event is $\beta_{4,b}(1-P_{\mathrm{LH}}')P_{0\mathrm{L}}$ and the expected BER in a packet of this event is denoted by  $\overline{\mathsf{BER}}_{4,2}$. According to the similar analysis in Appendix B, $\overline{\mathsf{BER}}_{4,2}\approx1/6$ and $\beta_{4,b}$ can be expressed as
\begin{equation}
\begin{split}
&\beta_{4,b}=\frac{\mathcal{OR}{\scriptsize\left|\begin{array}{ccc}
0, & 1, & \lfloor KL_s\rceil-K\\
\mathrm{A}_{\mathrm{L}},&K-1, &K-1 \\
\end{array}\right.}-\mathcal{OR}{\scriptsize\left|\begin{array}{ccc}
0, & 2, & \lfloor KL_s\rceil-K\\
\mathrm{A}_{\mathrm{L}},&K-2, &K-1 \\
\end{array}\right.}}{\mathcal{OR}{\scriptsize\left|\begin{array}{ccc}
0, & 1, & \lfloor KL_s\rceil-K\\
\mathrm{A}_{\mathrm{L}},&K-1, &K-1 \\
\end{array}\right.}},\\
\end{split}
\end{equation}

Event 3: The first OTD phase is wrongly detected, the second OSD phase is correctly detected and there are more than one chip error in the first OTD phase. The probability  of this event is $(1-\beta_{4,a})P_{\mathrm{LH}}'(1-P_{0\mathrm{L}})$ and the expected BER in a packet of this event is denoted by  $\overline{\mathsf{BER}}_{4,3}$, which is less than $1/2$ (the BER of random guess). 

Event 4: The first OTD phase is correctly detected, the second OSD phase is wrongly detected and there are more than one pair of chip errors in the second OSD phase. The probability  of this event is $(1-\beta_{4,b})(1-P_{\mathrm{LH}}')P_{0\mathrm{L}}$ and the expected BER in a packet of this event is denoted by  $\overline{\mathsf{BER}}_{4,4}$, which is less than $1/2$ (the BER of random guess). 

Event 5: Both OTD and OSD phases are wrongly detected. The probability  of this event is $P_{\mathrm{LH}}'P_{0\mathrm{L}}$ and the expected BER in a packet of this event is denoted by  $\overline{\mathsf{BER}}_{4,5}$, which is less than $1/2$ (the BER of random guess). 

Summarizing all above and substituting $P_{\mathrm{LH}}'$ in (\ref{PLHN}) and, $P_{0\mathrm{L}}$ in (\ref{P0L}), Theorem 4 can be derived.


\begin{IEEEbiography}[{\includegraphics[width=1in,height=1.25in,clip,keepaspectratio]{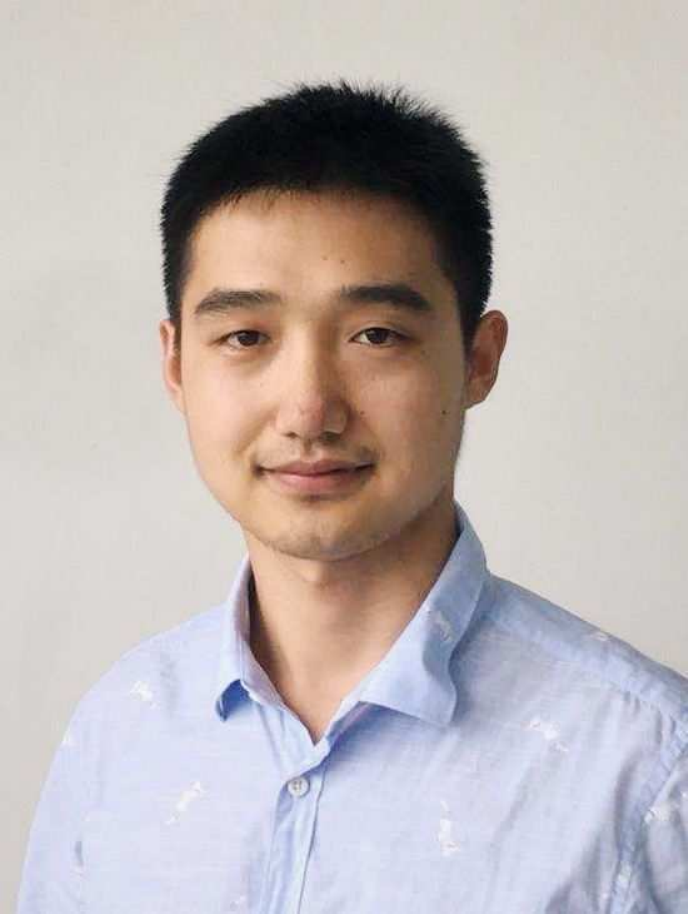}}]{Shuaishuai Guo}(S'14-M'17) received the B.E and Ph.D. degrees in communication and
information systems from the School of Information
Science and Engineering, Shandong University,
Jinan, China, in 2011 and 2017, respectively. He visited University of Tennessee at Chattanooga (UTC), USA, from 2016 to 2017. Now, he is working as a postdoctoral research fellow at King Abdullah University of Science and Technology (KAUST), Saudi Arabia, since 2017.

His research interests include wireless multiple-input multiple-output
communications, optical wireless communication and mobile edge computing.
\end{IEEEbiography}

\begin{IEEEbiography}[{\includegraphics[width=1in,height=1.25in,clip,keepaspectratio]{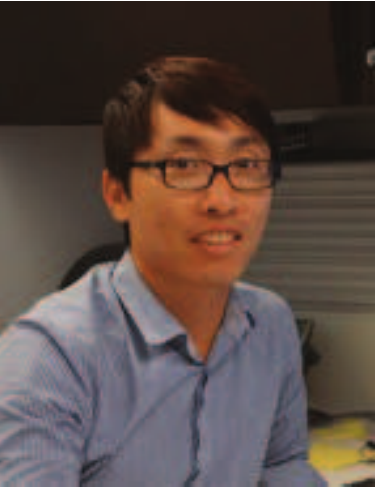}}]{Ki-Hong Park}(S'06-M'11)  received the B.Sc. degree in electrical, electronic, and radio engineering from Korea University, Seoul, Korea, in 2005 and the joint M.S. and Ph.D. degrees from the School of Electrical Engineering, Korea University, Seoul, Korea, in 2011. He joined KAUST as a postdoctoral fellow in April, 2011. Since December 2014, he has been working as a Research Scientist of Electrical Engineering in the Division of Computer, Electrical, Mathematical Science and Engineering (CEMSE), King Abdullah University of Science and Technology (KAUST), Thuwal, Saudi Arabia. His research interests are in the broad field of communication theory and its application to the design and performance evaluation of wireless communication systems and networks. On-going research includes the application to underwater visible light communication, optical wireless communications, unmanned aerial vehicle communication and physical layer secrecy.
\end{IEEEbiography}

\begin{IEEEbiography}[{\includegraphics[width=1in,height=1.25in,clip,keepaspectratio]{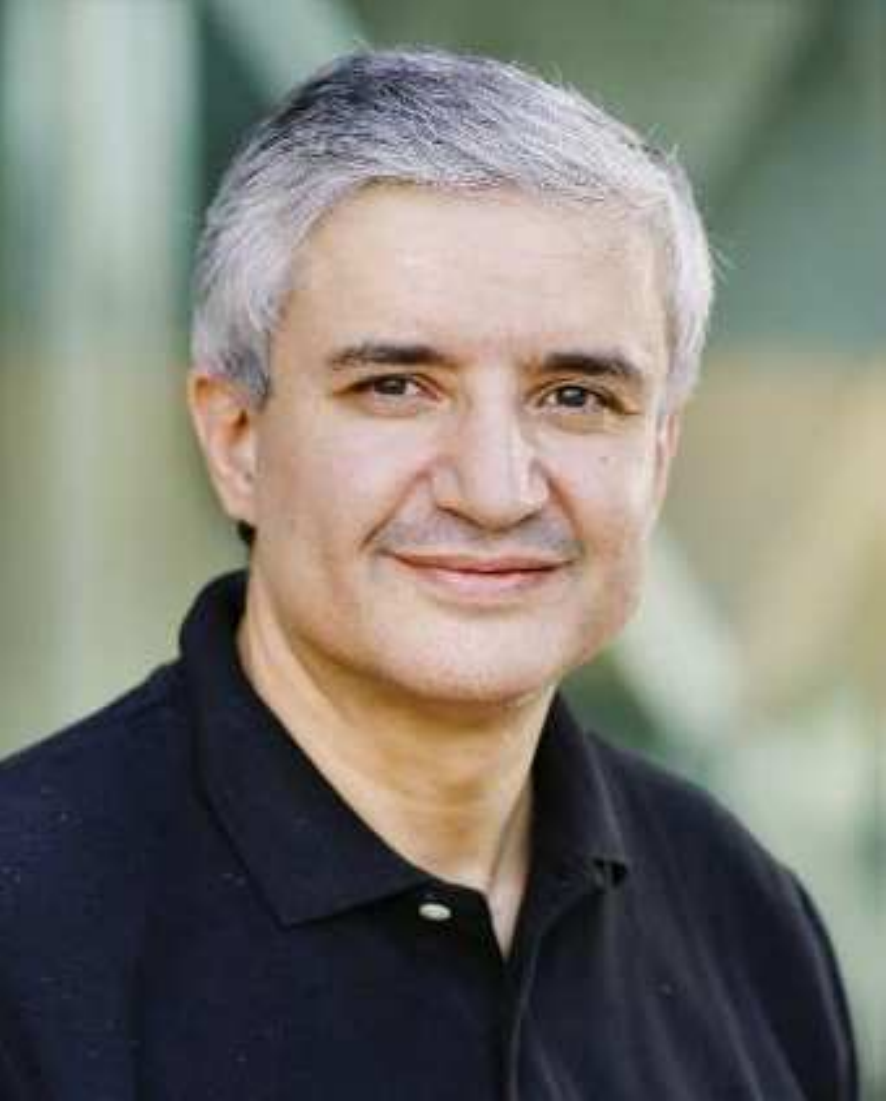}}]{Mohamed-Slim Alouini} 
(S'94-M'98-SM'03-F'09)  was
born in Tunis, Tunisia. He received the Ph.D. degree in Electrical Engineering
from the California Institute of Technology (Caltech), Pasadena,
CA, USA, in 1998. He served as a faculty member in the University of Minnesota,
Minneapolis, MN, USA, then in the Texas A\&M University at Qatar,
Education City, Doha, Qatar before joining King Abdullah University of
Science and Technology (KAUST), Thuwal, Makkah Province, Saudi
Arabia as a Professor of Electrical Engineering in 2009. His current
research interests include the modeling, design, and
performance analysis of wireless communication systems.
\end{IEEEbiography}

\end{document}